\documentclass[aip, cha, reprint, unsortedaddress]{revtex4-1}
\usepackage[ansinew]{inputenc}
\usepackage{amsmath}
\usepackage{amsfonts}
\usepackage{amssymb}
\usepackage{graphicx}
\usepackage{hyperref}
\begin{document}
\newcommand{\D}{\mathrm{d}}
\newcommand{\I}{i}
\newcommand{\E}{\mathrm{e}}
\newcommand{\td}[2]{\frac{\D #1}{\D #2}}
\newcommand{\pd}[2]{\frac{\partial #1}{\partial #2}}
\newcommand{\tr}{\mathop{\mathrm{tr}}}
\newcommand{\arcsinh}{\mathop{\mathrm{arcsinh}}}
\newcommand{\arccosh}{\mathop{\mathrm{arccosh}}}
\newcommand{\Ai}{\mathop{\mathrm{Ai}}}
\newcommand{\Bi}{\mathop{\mathrm{Bi}}}
\newcommand{\sign}{\mathop{\mathrm{sgn}}}
\newcommand{\order}{\mathcal{O}}
\newcommand{\thetas}{\tilde\theta}
\newcommand{\Fq}[1]{\Phi_{#1}}
\newcommand{\Fqi}[2]{\Phi_{#1, #2}}
\newcommand{\Pq}[1]{P_{#1}}
\newcommand{\Pqi}[2]{P_{#1, #2}}
\newcommand{\Pt}[1]{{\tilde P}_{#1}}
\newcommand{\Pti}[2]{{\tilde P}_{#1, #2}}
\newcommand{\thetap}[1]{\theta_{#1}}
\newcommand{\PMap}{\Pi}
\newcommand{\Pfp}{\theta^*}
\newcommand{\cov}{\mathcal{C}}
\newcommand{\EiE}[2]{E\left(#1\vphantom{#2} \right|\left. \vphantom{#1} #2 \right)}
\newcommand{\avgcos}{\langle \cos \theta \rangle}
\renewcommand{\Re}{{\mathop{\mathrm{Re}}}}
\renewcommand{\Im}{{\mathop{\mathrm{Im}}}}
\newcommand{\Ct}{{\sigma_1}}
\newcommand{\Cs}{{\sigma_3}}
\newcommand{\Cx}{{\sigma_2}}
\newcommand{\Mea}{{e_1}}
\newcommand{\Meb}{{\tilde e}}
\newcommand{\tstar}{{t^{*}}}
\newcommand{\Mred}{{\tilde M}}
\newcommand{\eps}{\varepsilon}
\newcommand{\BZ}{B_0}
\newcommand{\weq}{w_\text{eq}}
\newcommand{\thetab}{{\overline{\theta}}}
\newcommand{\thetat}{{\tilde \theta}}
\newenvironment{Mtx}{\left(\begin{array}{cc}}{\end{array}\right)}
\newenvironment{Cvec}{\left(\begin{array}{c}}{\end{array}\right)}
\newcommand{\qvec}[2]{\left(\begin{array}{c} #1 \\ #2 \end{array}\right)}
\newcommand{\ie}{\textit{i.e.}}
\newcommand{\eg}{e.g.}
\newcommand{\cf}{cf.}
%
%
\title{Stability properties of periodically driven overdamped pendula
  and their implications to physics of semiconductor superlattices and
  Josephson junctions}
\author{Jukka Isoh\"{a}t\"{a}l\"{a}}
\affiliation{Department of Physical Sciences, P.O. Box 3000, University of Oulu FI-90014, Finland}
\author{Kirill N. Alekseev}
\affiliation{Department of Physics, Loughborough University LE11 3TU, United Kingdom} 
\affiliation{Department of Physical Sciences, P.O. Box 3000, University of Oulu FI-90014, Finland}
\date{\today}
\begin{abstract}
  We consider the first order differential equation with a sinusoidal
  nonlinearity and periodic time dependence, that is, the periodically
  driven overdamped pendulum. The problem is studied in the case that
  the explicit time-dependence has symmetries common to pure ac-driven
  systems.  The only bifurcation that exists in the system is a
  degenerate pitchfork bifurcation, which describes
  an exchange of stability between two symmetric nonlinear modes.
  Using a type of Pr\"{u}fer transform to a pair of linear
  differential equations, we derive an approximate condition of the
  bifurcation.  This approximation is in very good agreement with our
  numerical data. In particular, it works well in the limit of large
  drive amplitudes and low external frequencies.  We demonstrate the
  usefulness of the theory applying it to the models of pure ac-driven
  semiconductor superlattices and Josephson junctions.
  We show how the knowledge of bifurcations in the
  overdamped pendulum model can be utilized to describe effects of
  rectification and amplification of electric fields in these
  microstructures.
\end{abstract}
\maketitle
%
%

\begin{quotation}
Pendulum and pendulum-like equations are arguably among the most
important classes of equations in modern nonlinear
science\cite{sagdeev}. The most often encountered representatives of
this family may well be the driven and damped pendulum, $\ddot \theta
+ \gamma \dot \theta + \sin \theta = f(t)$, and its first order
counterpart, the overdamped pendulum, $\gamma \dot \theta + \sin
\theta = f(t)$, the latter type being the topic of this paper. These
equations appear, for instance, in the well-known
Stewart-McCumber\cite{mccumber, stewrt} and
Aslamazov-Larkin\cite{aslamazov69} models of Josephson junctions.  The
sinusoidal nonlinearity gives rise to a wide class of nonlinear
phenomena that have important practical applications: the ac-Josephson
effect\cite{langenberg, levinsen} and the modern voltage
standard\cite{hamilton} are prime examples of this. More recently,
pendulum equations have been found in the theory of semiconductor
superlattices where they frequently occur in the limiting cases of the
governing differential equations\cite{alekseev98:spont-dc, cannon01,
  alekseev02a, dodin98b, alekseev05:lsslcr, zharov06, dodin03}.
Moreover, overdamped pendulum equations are often
encountered in mathematical models of synchronization of nonlinear
oscillators\cite{pikovsky01}.  Further recent interest in the
overdamped pendula has come from the field of high-$T_c$
superconductors: It has been demonstrated that stacked array of
intrinsic Josephson junctions in  magnetic field can
be synchronized and described by overdamped pendulum-like
dynamics\cite{gaifullin08, gaifullin09}.  Properties of ac-driven
overdamped pendula are also of importance in theories of the
amplification of microwave radiation in Josephson point
contacts\cite{kuzmin79, kuzmin80, *kuzmin80-orig,
  velichko99,likharev-c10-11}.  And last but not least, in our
previous work\cite{isohatala05:sbpend} we demonstrated how
instabilities occurring in the overdamped pendulum are carried over to
higher dimensional systems such as the strongly damped second order
pendulum equation. Our present paper has two sides: mathematical and
physical. Here we develop a mathematical technique which allows to
find bifurcations in a class of overdamped pendula models for a wide
range of their parameters, including a difficult but physically
interesting case of low frequencies of driving force. We also show how
this technique can be applied to symmetric physical systems
demonstrating pendulum dynamics in some limiting cases.  Our main
focus is on the rectification and amplification of microwave radiation
in unbiased semiconductor superlattices and Josephson junctions.
\end{quotation}

\section{\label{sec:intro}Introduction}
We consider the first order ordinary differential equation with a
sinusoidal nonlinearity and arbitrary time dependence
\begin{equation}
  \dot \theta(t) + G(t) \sin \theta(t) = F(t).
  \label{eq:gen}
\end{equation}
The dynamics of the overdamped pendulum has been
studied by several people, motivated by direct physical applications
mentioned above. In spite of its apparent simplicity, novel nonlinear
dynamics, \eg\ strange nonchaotic attractors\cite{romeiras87} have
been found. Here, we restrict ourselves to a specific class of
periodic forcing, specifically consider bifurcations occuring in this
systems, and its applications in physical systems.

We will implicitly assume everywhere that $F$ and $G$ are real,
continuous, and differentiable sufficiently many times.  Our focus
will be on functions $F$ and $G$ that have the following property
\begin{equation}
  F(t + T/2) = -F(t), \qquad G(t + T/2) = G(t).
  \label{eq:Ssymm}
\end{equation}
With the above choice of external time-dependence, Eq.~(\ref{eq:gen})
remains invariant under the transformation
\begin{equation}
  t \to t + T/2, \quad 
  \theta(t) \to -\theta(t + T/2) + 2k\pi, 
  \label{eq:Qsymm}
\end{equation}
and $k$ is an integer.  This type of forcing and the associated
symmetry are of interest in many pure ac-driven physical systems, in
particular bulk semiconductors and semiconductor superlattices,
where breaking of symmetry (\ref{eq:Qsymm}) implies generation of a
spontaneous dc bias\cite{bumyalene89, alekseev98:spont-dc}.  In our
previous work\cite{isohatala05:sbpend} we in passing considered
Eq.~(\ref{eq:gen}) with $G(t) = 1$ and $F(t) = f \cos \omega t$. We
observed that the only instability that occurs is an exchange of
stability between  two periodic solutions following
symmetry~(\ref{eq:Qsymm}) and having the properties $\langle \theta
\rangle = 0$ and $\langle \theta \rangle = \pi$, where $\langle \cdot
\rangle$ stands for time-average across the period of the solution.

We found that the symmetry breaking bifurcation,
($\langle \theta \rangle \neq 0, \pi$) \cite{dhumieres82, swift84} of
the strongly damped second order pendulum
\begin{equation}
  \ddot \theta + \gamma \dot \theta + \sin \theta = f \cos \omega t,
  \label{eq:ddp}
\end{equation}
reduced to this instability in the limit of very large damping.  We
conjectured that other ac-driven systems reducible to
Eq.~(\ref{eq:gen}) undergo a type of bifurcation similar to the one
found in the strongly damped pendulum equation near the points where
the exchange of stability occurs.  This motivates our present extended
study of the stability properties of system~(\ref{eq:gen}) and
applications to a number of physical systems.

In this paper, our mathematical analysis is based on mapping of
equation~(\ref{eq:gen}) to a particular second order linear
differential equation.  Such transformations have proved useful in the
study of various linear and nonlinear differential
equations. Following Pr\"{u}fer's application of the idea to
Sturm-Liouville problems\cite{prufer26}, these changes of variables
are sometimes called Pr\"{u}fer transforms. In a sense the reverse of
this approach was taken in by Bondeson \textit{et
  al.}\cite{bondeson85:quasi-pend} where the authors used a similar
transformation to study quasiperiodically driven overdamped equation
by relating the problem to a Schr\"{o}dinger equation with a
quasiperiodic potential. We take essentially the same approach, but
focus on the more specific problem of periodically driven equation.

On the other hand, applications considered in this paper are based on
the connection of an exchange of stability in the pendulum with the
physical phenomena of amplification and rectification.  Here we
consider effects of microwave rectification and amplification in two
pure ac-driven systems reducible to the overdamped pendulum:
single-band lateral semiconductor superlattice and point-contact
Josephson junction.

 Formally, by rectification we mean the
conversion of pure ac excitation into response at even harmonics of
some quantity that is a odd function of $\theta$; for instance
$\theta$ itself or $\sin \theta$. As an example of rectification,
consider Eq.~(\ref{eq:ddp}) as a toy model where the drive $f(t)$
corresponds to some ac applied field and that current is $j(t) \propto
\sin \theta$. Rectification would then imply $j_\text{dc} = \langle
j(t) \rangle \neq 0$, \ie\ obtaining a direct current response from a
pure ac excitation, hence the term ``rectification''. For symmetric
solutions $\langle j(t) \rangle = 0$ rectification is impossible,
since $\theta$ will only have odd harmonics, excluding possibly zeroth
harmonic that is a multiple of $\pi$. Thus, symmetry breaking is
a prerequisite for rectification.

Note that rectification due to spontaneous breaking of symmetry in
solutions, Eq.~(\ref{eq:Qsymm}), should be distinguished from
phase-dependent rectification due to breaking of symmetry in the
equations\cite{flach00}. The latter requires explicitly introducing
$f$ that does not follow Eq.~(\ref{eq:Ssymm}), for example an
additional phase-shifted second harmonic $\cos(\omega t)+\cos(2\omega
t+\phi)$.

For large damping we get overdamped first
order pendulum for which exchange of stability arises for same
parameters as symmetry breaking in second order
pendulum. Rectification in pendulum is, however, rather artificial
model which does not correspond to any real physical system.
Nevertheless, in dynamical systems describing realistic physical
situations, symmetry breaking bifurcation is realized near values of
parameters that are close to the values necessary for the exchange of
stability in the overdamped pendulum.

Here we consider a model of ac-driven lateral
semiconductor superlattice\cite{dodin98b, alekseev05:lsslcr} which is
described by two first-order nonlinear balance equations which can be
reduced to a sort of overdamped pendulum (see Eq.~(\ref{eq:latpend})
in the limit of strong nonlinearity. Symmetry breaking in balance
equations of lateral superlattice corresponds to rectification of
applied ac electric field \cite{alekseev05:lsslcr}. We demonstrate how
analysis of instabilities in Eq.~(\ref{eq:gen}) can provide a quite
useful information on the parameter space of rectification in these
nanostructures.

Our another application is related to
amplification of infinitesimally weak signal in Josephson point
contact described by Eq.~(\ref{eq:gen}), in which $\dot{\theta}$ --
voltage, $f(t)$ -- current and $\theta$ itself is difference of phases
of wave functions of superconductors in the junction. As a rule, an
amplification of a small signal is observed near the onset of a
dynamical instability\cite{wiesenfeld85, wiesenfeld86}. Here we show a
small-signal amplification near an exchange of stability and at
frequencies of signal close to even harmonics of pump. Therefore, this
effect of amplification can be considered also reminiscent of symmetry
breaking bifurcation in strongly damped second order pendulum
Eq.~(\ref{eq:ddp}).  Despite here even harmonics are forbidden by
symmetry, nontrivial amplification of additional weak signal does
exist at pump amplitudes and frequencies close to those necessary to
realize real symmetry breaking.

The outline of this paper is as follows.  In the next section we will
introduce the change of variables that yields a second order linear
differential equation and briefly recapitulate on some known
properties of its solution and their implications on
Eq.~(\ref{eq:gen}).  We will then proceed to the more specific problem
of forcing following Eq.~(\ref{eq:Ssymm}) and show that an exchange of
stability is the only instability occuring in this system.  We then
consider perturbations of the Eq.~(\ref{eq:gen}) and essentially prove
our earlier conjecture that the exchange of stability is a limit of
pitchfork bifurcations.  In the subsequent section, we shift to a more
practical approach: An approximate condition for the instability to
occur will be derived in nontrivial case of large $F$ and
$G$. Finally, we go on to apply the results to relevant physical
problems. Technical details are presented in three Appendixes.

\section{Equivalent linear equation}
We start by introducing a change of variables from $\theta(t)$ to new
variables $q_1(t), q_2(t)$ as
\begin{equation}
  \theta(t) = 2 \arctan\left(\frac{q_1(t)}{q_2(t)}\right).
  \label{eq:cov}
\end{equation}
We will denote the vector $(q_1(t), q_2(t))^T$ by $Q$ and the change
of variables by $\theta(t) = \cov[Q(t)]$. Since Eq.~(\ref{eq:cov})
alone does not fix the functions $q_1$, $q_2$, we have some freedom in
choosing the differential equations for the new variables. Here we opt
for a particularly symmetric form of the equations
\begin{equation}
  \frac{\D}{\D t}
  \begin{Cvec} q_1(t) \\ q_2(t) \end{Cvec}
  =
  \frac{1}{2}
  \begin{Mtx}
    -G(t) & F(t) \\
    -F(t) & G(t) 
  \end{Mtx}
  \begin{Cvec} q_1(t) \\ q_2(t) \end{Cvec}.
  \label{eq:lin}
\end{equation}
The coefficient matrix on the right-hand side of Eq.~(\ref{eq:lin})
will be denoted by $A$.  We consider only periodic $F$ and $G$, and
therefore Floquet theory can be directly applied to the problem.  We
adopt the following notations for the Floquet solutions $\Fq{i}$, $i =
1, 2$:
\begin{equation}
  \Fq{i}(t) 
  = \E^{B_i t} \Pq{i}(t) = \E^{\Re\{B_i \} t} \Pt{i}(t),
\end{equation}
where $B_i$ are the (complex) characteristic exponents and $\Pq{i}(t)$
are $T$-periodic functions. Functions $\Pt{i}(t)$ contain the
oscillating parts of the Floquet solutions.  Additional lower indices
will label the component of $\Fq{i}$, $\Pq{i}$, and $\Pt{i}$,
\eg\ $\Fq{i} = (\Fqi{i}{1}, \Fqi{i}{2})^T$.  Due to the vanishing
trace of $A$, the characteristic exponents have always the form (a)
$B_{1,2} = \pm \BZ + 2 \pi \I k_{1,2}/T$, where $\BZ$ is real and
$k_{1,2}$ are integers, or (b) $B_{1,2} = 2 \pi \I (\pm r +
k_{1,2})/T$ where $r$ is real and not an integer.  In the next
section, we will show that the latter case is never realized if
symmetry~(\ref{eq:Ssymm}) applies, and thus case (b) will not be
addressed in what follows.  However, the special case of $B_{1,2} = 2
\pi \I k_{1,2} / T$ ($\BZ = 0$) will be covered.

The Floquet solutions give a complete description of dynamics of
$\theta(t)$. Supposing case (a) from above holds, the general
solution of the pendulum equation is
\begin{equation}
  \theta(t) 
  = \cov\left[\cos \frac{\psi_0}{2} \E^{\BZ t}\Pt{1} 
    + \sin \frac{\psi_0}{2} \E^{-\BZ t}\Pt{2}\right],
  \label{eq:gensol}
\end{equation}
where $\psi_0$, is a constant that depends on the initial value of
$\theta$, and whose value over $-\pi < \psi \leq \pi$ uniquely
determines the solution $\theta$ up to modulo $2\pi$.  Clearly, when
$\BZ \neq 0$ there are exactly two periodic solutions $\thetap{i}$, $i
= 1, 2$, that are simply given by the Floquet solutions:
\begin{equation}
  \thetap{i}(t) = 
  2\arctan
  \left( 
  \frac{\Fqi{i}{1}(t)}{\Fqi{i}{2}(t)}
  \right).
  \label{eq:thetap}
\end{equation}
From Eq.~(\ref{eq:gensol}) it follows that stable solutions of the
overdamped pendulum correspond to the unstable solutions of the linear
equation and vice-versa. This can be also seen from the relation that
appears in Refs.~\onlinecite{johnson82, bondeson85:quasi-pend} and
also applies here
\begin{equation}
  -\Lambda = \frac{1}{T} \int_0^{T} G(t') \cos \theta (t') \; \D t' = 2 |\BZ|,
  \label{eq:avgcos}
\end{equation}
where $\Lambda$ is the average exponential rate of growth for a
infinitesimal perturbation of $\theta$. Negative of its absolute value
coincides with the maximal Lyapunov exponent of Eq.~(\ref{eq:gen}).
Periodicity of the asymptotic solutions in the sense that $\theta(t +
T) = \theta(t) + 2\pi n$ also follows immediately from the form of the
Floquet solutions.

\section{\label{sec:symm}Symmetric case}
We now turn to our findings regarding Eq.~(\ref{eq:lin}) with forcing
following Eq.~(\ref{eq:Ssymm}). Matrix $A$ transforms in the
$T/2$-shift as $A(t + T/2) = \Meb A(t) \Meb$, where
\begin{equation}
  \Meb = \begin{Mtx} 1 &   0 \\  0 & -1 \end{Mtx}. \label{eq:xmats}
\end{equation}
This property enables us to write the principal matrix $U$, $\dot U =
A U$, $U(0) = I$, on the latter half of a drive cycle in terms of the
former
\begin{equation}
  U(t + T/2) = \Meb U(t) \Meb U(T/2).
  \label{eq:Ushift}
\end{equation}
The monodromy matrix $M = U(T)$ can now be factored into a square of
the matrix $\Mred = \Meb U(T/2)$, and the eigenvalue equation
determining the Floquet solutions, $M \Fq{i}(0) = \exp(B_iT)
\Fq{i}(0)$, can be solved using the matrix $\Mred$ instead of
$M$. Determinant of $\Mred$ equals $-1$, and thus the solution to the
characteristic equation becomes
\begin{equation}
  B_{1,2} = 
  \left\{
  \begin{array}{l} 
    \phantom{-}\frac{2}{T} \arcsinh \left(\frac{1}{2} \tr \Mred\right), \\
    -\frac{2}{T} \arcsinh \left(\frac{1}{2}\tr \Mred\right) + \I \frac{2\pi}{T}.
  \end{array}
  \right.
  \label{eq:bsymm}
\end{equation}
This shows $B_{1,2}$ are always real or real plus an integer multiple
of $2\I \pi/T$. We will make frequent use of real part of $B_1$,
$\Re\{B_1\} = \BZ = 2 \arcsinh(\tr \Mred/2)/T$. 

For the periodic parts $\Pq{i}$ of the Floquet solutions, the
following now holds. Applying Eq.~(\ref{eq:Ushift}) and $\Mred
\Fq{i}(0) = \exp(B_iT/2) \Fq{i}(0)$ one finds that $\Pq{i}(t + T/2) =
\Meb \Pq{i}(t)$. Component-wise this property reads:
\begin{subequations}
  \begin{eqnarray}
    \Pqi{i}{1}(t + T/2) &=& \Pqi{i}{1}(t), \\
    \Pqi{i}{2}(t + T/2) &=& -\Pqi{i}{2}(t).
  \end{eqnarray}
  \label{eq:pper}
\end{subequations}
That is, the first component of $\Pq{i}$ is $T/2$-periodic, and the
second $T/2$-antiperiodic.  It immediately follows that the periodic
solutions $\thetap{i}$ of Eq.~(\ref{eq:gen}) are symmetric in the
sense of Eq.~(\ref{eq:Qsymm}).  Note that our definition for $B_2$,
Eq.~(\ref{eq:bsymm}), includes an imaginary component, which
contributes to the oscillating part of $\Fq{2}$.  Therefore, $\Pq{2}$
is not real, and it is then more convenient to use the functions
$\Pt{i}$ instead. Now $\Pt{1} = \Pq{1}$, and so $\Pt{1}$ has the same
periodicity as $\Pq{1}$. On the other hand $\Pt{2} = \exp(2 \I \pi t /
T) \Pq{2}$, and thus $\Pti{2}{1}$ is $T/2$-antiperiodic and
$\Pti{2}{2}$ is $T/2$-periodic.

Eq.~(\ref{eq:pper}) also determines two properties regarding the
rotations and the average value of the periodic solutions
$\thetap{i}$.  Here we assume that roots of $F$ are simple, that is,
if $F(t) = 0$ then $\dot F(t) \neq 0$. We aim to connect the number of
zeros of $\Fqi{i}{2}(t)$ over $0 \leq t < T/2$, here denoted $n$, to
physically relevant properties of $\thetap{i}$ -- it will be shown
that $n$ indeed has significance to dynamics of real physical systems
we are considering. Since $\thetap{i}$ is symmetric, we can write
$\thetap{i}(T/2) = -\thetap{i}(0) + 2 \pi j$, where $j$ counts the
positive direction crossings of the line $\theta = \pi \mod 2\pi$.
From Eq.~(\ref{eq:thetap}) it can be seen that these crossings occur
at simple zeros of $\Fqi{i}{2}$, and from Eq.~(\ref{eq:gen}) that the
direction of the crossing is given by the sign of $F$. Using again the
symmetry, the average of $\thetap{i}$ over $t = 0 \ldots T$, $\langle
\thetap{i} \rangle$, is $j \pi$. Now, it is easy to see that the
parities of $n$ and $j$ are the same,
and thus $\langle \thetap{i} \rangle = n \pi \mod 2\pi$.  From
Eq.~(\ref{eq:pper}) it is clear that $n$ is odd for $\thetap{1}$ and
even for $\thetap{2}$, and so $\langle \thetap{1} \rangle = \pi$ and
$\langle \thetap{2} \rangle = 0$, both modulo $2\pi$. This shows that
our previous finding\cite{isohatala05:sbpend} regarding the averages
of $\theta$ regarding the case $F(t) = f \sin \omega t$ and $G(t) = 1$
holds in general.

Further, $n$ relates to the oscillations of $\sin \thetap{i}$ and
other quantities that are periodic $\theta$. For instance, consider
$F$ is such that $F(t) > 0$ ($< 0$) for $0 < t < T/2$ ($T/2 < t <
T$). As $\theta$ rotates it passes the upright vertical position $n$
times and always in the positive direction, and so $n$ gives the
minimum and maximum number of oscillations of $\sin \thetap{i}$ or
$\cos \thetap{i}$ in one half drive period. This has implications to
the physical systems we are considering, since in these $\sin \theta$
and $\cos \theta$ have relevant physical interpretations.

\subsection{Exchange of stability}
Having established that the characteristic exponents are always real
plus integer multiples of $\I 2\pi / T$, it then follows that the only
possible type of instability is an exchange of stability where one
Floquet solution loses stability and the other gains it. This in turn
occurs when $\tr \Mred = 0$ and as consequence $\BZ = 0$. Noting that
eigenvalues of $\Mred$ are never equal, one finds that vectors
$\Fq{i}(t)$, $i = 1, 2$, are linearly independent for all $t$ and for
any $\BZ$. Thus, the Floquet solutions $\Fq{i}$ never map to same
solution of Eq.~(\ref{eq:gen}) and the corresponding asymptotic
solutions $\thetap{i}(t) = \cov[\Fq{i}(t)]$ never cross each other as
a parameter is varied.  Consequently, the stability is exchanged
without the solutions colliding, in contrast to a transcritical
bifurcation.  We will later show, however, that the exchange of
stability can be seen as a type of pitchfork bifurcation.  From
Eq.~(\ref{eq:gensol}) it is clear that when $\BZ = 0$ all solutions to
the overdamped pendulum are periodic. Since the superposition
$\cos(\psi_0/2) \Pt{1} + \sin(\psi_0/2) \Pt{2}$ that gives the general
solution, Eq.~(\ref{eq:gensol}), does not have periodicity analogous
to Eq.~(\ref{eq:pper}), the corresponding solutions $\theta$ are not
symmetric.

Further, crossing the instability has a clear effect on some relevant
quantities. As was discussed above, the number of simple roots of
$\Fqi{i}{2}$ relates to the rotations and the average of $\theta$.
Let then $\Fq{+}$ be the unstable Floquet solution, $\thetap{+} =
\cov[\Fq{+}]$ the stable periodic solution of the overdamped pendulum,
and $n$ the number of simple roots of $\Fq{+}(t)$ over $0 \leq t <
T/2$. Because at the instability $\Fq{+}$ switches between being
$\Fq{1}$ and $\Fq{2}$, $n$ changes by one. Consequently, the average
value of the stable periodic solution jumps by $\pi$, and further,
since the value of $n$ is intimately connected to the oscillations of
$\sin \theta$ and $\cos \theta$, these quantities exhibit a change in
in their frequency spectrum. We will later apply this finding in the
section on lateral semiconductor superlattices. Clearly, the integer
$n$ partitions the parameter space into disjoint regions with the
instability separating them. Thus, $n$ serves as a convenient label
for different regions of parameter space.

\begin{figure}
  \includegraphics{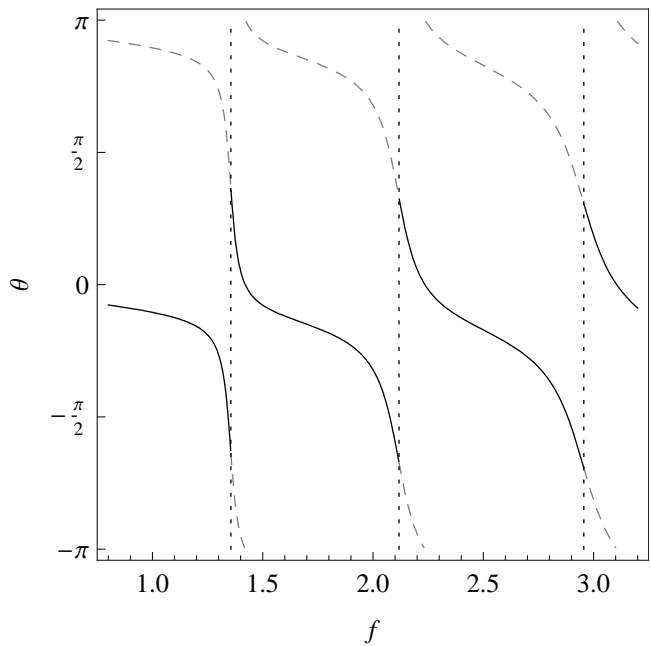}
  \caption{\label{fig:bifd} Bifurcation diagram showing the fixed
    points of the stroboscopic map $\PMap$ of Eq.~(\ref{eq:gen}) as a
    parameter is varied. Here $G(t) = 1$ and $F(t) = f \sin \omega t$,
    where $\omega = 0.3$. Forcing amplitude $f$ is taken as the
    control parameter. Solid and dashed lines indicate stable and
    unstable fixed points respectively, while dotted line indicates
    marginally stable fixed points. These span the whole phase space
    and occur exactly at $\BZ = 0$.}
\end{figure}

\begin{figure}
  \includegraphics{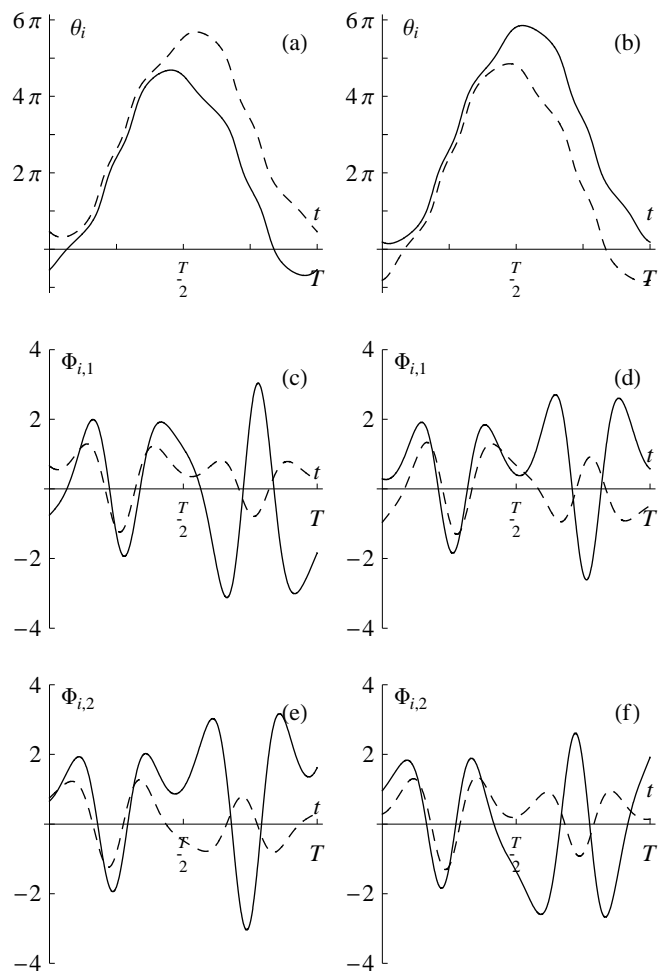}
  \caption{\label{fig:floq} Representative periodic solutions
    $\thetap{i}$, $i=1, 2$, of Eq.~(\ref{eq:gen}) with $F(t) = f \sin
    \omega t$, $G(t) = 1$ and the corresponding Floquet solutions
    $\Fq{i}$ just before [subfigures (a), (c), and (e)] and after a
    bifurcation [subfigures (b), (d), and (f)]. In both (a) and (b),
    the stable (unstable) periodic solution $\theta$ is plotted with a
    solid (dashed) line. Similarly, Floquet solution mapping to the
    stable (unstable) $\theta$ is plotted with solid (dashed)
    line. Note that it is the unstable $\Fq{}$ that corresponds to the
    stable $\theta$.  In the left-hand side subfigures (a, c, e),
    parameter $f = 2.9$ which is just below the critical value of $f =
    2.956$, while on the right-hand side (b, d, f) the parameter $f$
    is just above it, $f = 3$.}
\end{figure}

To help illustrate the bifurcation, we
introduce a Poincaré map $\PMap$. Naturally we take this to be the
stroboscopic map, defined so that $\theta(t_0 + T) =
\PMap(\theta(t_0))$ for some fixed $t_0$ which we take to be zero. If
$\BZ \neq 0$, the two fixed points $\Pfp_{i}$, $i = 1, 2$, of $\PMap$
are given in terms of the eigenvectors of $\Mred$, or Floquet
solutions at $t=0$, $\Fq{i}(0)$:
\begin{equation}
  \Pfp_i = 2\arctan\left(\frac{\Fqi{i}{1}(0)}{\Fqi{i}{2}(0)}\right).
  \label{eq:pfp}
\end{equation}
From the discussion above it follows that at $\BZ = 0$, $\PMap(\theta)
= \theta$ for all $\theta$. In Fig.~\ref{fig:bifd} we have plotted a
representative bifurcation diagram.  We take $G(t) = 1$ and $F(t) = f
\sin \omega t$, where $\omega = 0.3$, and plot the fixed points
$\Pfp_{i}$ as a functions of the forcing amplitude $f$. From the
diagram the bifurcation scenario can be easily visualized. At $f = 0$
(outside the plot range) we have two fixed points: $\theta = 0$,
$\pi$, where the latter is naturally the unstable point, since it
corresponds to the upright position of the pendulum.  At $f \simeq
1.356$ we find the first bifurcation.  The initially unstable state
becomes the stable one and vice-versa as the critical $f$ is crossed.
Exactly at the bifurcation, every point $\theta$ is a fixed point of
the Poincaré map. From there on, increasing $f$ further, we find the
the fixed points exchange their stabilities again at $f \simeq 2.118$
and $f \simeq 2.956$, with the marginally stable points spanning the
whole phase space exactly at the bifurcation.

The bifurcation described above bears
resemblance to a pitchfork bifurcation. In fact it can be seen as a
degenerate pitchfork bifurcation (PB), since two new branches of fixed
points emerge at the critical point. By degeneracy, we mean that these
branches exist only exactly at the bifurcation point, span the whole
phase space, and and are thus only marginally stable.  This is in
contrast to the (non-degenerate) PB, where the two additional branches
of fixed points exist before or after the bifurcation, that are either
stable (supercritical case) or unstable (subcritical case).  We note
that this is reminiscent of the scenario observed for the second
order, strongly damped pendulum\cite{isohatala05:sbpend}, where a
(non-degenerate) pitchfork bifurcation was found near the criterion
for exchange of stability in the overdamped equation.

In terms of normal forms, non-degeneracy is understood as
non-vanishing of a number of higher order derivatives with respect to
the variable of the flow at an equilibrium\cite{kuznetsov98}. We will
see that this is indeed the case when we map Eq.~(\ref{eq:gen}) to an
autonomous equation that is effectively a normal form on the circle
$0\ldots 2\pi$.  The form of the general solution,
Eq.~(\ref{eq:gensol}) suggests a natural way of mapping
Eq.~(\ref{eq:gen}) into an equivalent autonomous form: the periodic
parts $\Pt{i}$ describe the rotations of $\theta$, while the exchange
of stability is determined by the exponential factors and the
superposition phase $\psi_0$. It seems natural to use a trial function
where the constant phase $\psi_0$ is replaced by a time-dependent
$\psi(t)$, that also accounts for the exponential factors:
\begin{equation}
  \theta(t) = 2 \arctan \left(
  \frac{\cos \frac{\psi(t)}{2} \Pti{1}{1}(t) + \sin \frac{\psi(t)}{2} \Pti{2}{1}(t)}
       {\cos \frac{\psi(t)}{2} \Pti{1}{2}(t) + \sin \frac{\psi(t)}{2} \Pti{2}{2}(t)}
  \right).
  \label{eq:slowtrial}
\end{equation}
We substitute this into Eq.~(\ref{eq:gen}) and use Eq.~(\ref{eq:lin})
to obtain an equation for $\psi$:
\begin{equation}
  \dot \psi = -2\BZ \sin \psi.
  \label{eq:slowphase}
\end{equation}
No approximations were needed to derive Eq.~(\ref{eq:slowphase}). This
equation describes the approach to limit-cycles for the whole class of
systems. Obviously, it is also the simplest non-trivial overdamped
pendulum that has symmetry~(\ref{eq:Ssymm}). Eq.~(\ref{eq:slowphase})
has equilibria $0$, $\pi$ when $\BZ \neq 0$, in the case $\BZ = 0$
every point is an equilibrium, \cf\ Fig.~\ref{fig:bifd}.  To compare
bifurcations of Eq.~(\ref{eq:slowphase}) to the pitchfork bifurcation,
we recall that normal form of a PB is $\dot r = r (\mu \pm r^2)$,
where $\mu$ is the bifurcation parameter\cite{kuznetsov98}. From
Eq.~(\ref{eq:slowphase}), we see that near the equilibrium $0$
($\pi$), $\dot \psi = \mp \mu \psi (1 - \psi^2/6 + \cdots)$, where
$\mu = 2\BZ$. Thus, we see that whereas in the non-degenerate case,
only the leading order term vanishes at the bifurcation, here the
right-hand side becomes identically zero at $\mu = 0$.

 To help visualize how the Floquet solutions
relate to the solutions periodic solutions of the pendulum, we have
plotted in Fig.~\ref{fig:floq} the periodic solutions $\thetap{i}(t) =
\cov[\Fq{i}(t)]$, together with the Floquet solutions $\Fq{i}$.  We
have used the same $G$ and $F$ as above and take $f$ to be near the
third bifurcation shown in Fig.~\ref{fig:bifd}. The left-hand side
subfigures (a), (c), and (e) show the solutions for $f = 2.9$ which is
just smaller than the critical value of $f = 2.956$, while subfigures
(b), (d), and (f) are plotted for $f = 3.0$. From (a) and (b) it can
be seen that the $\theta_i$ change little across the bifurcation
point, only the stability is exchanged.  Similarly, the Floquet
solutions remain roughly unchanged as the as the critical $f$ is
crossed, excluding the fact that the exponential envelope switches
from decaying to diverging or vice-versa.

\subsection{\label{sec:symbr}Effect of perturbations}
From Eq.~(\ref{eq:slowphase}) it is evident that near a bifurcation
the system is structurally unstable. An important question is then how
dynamics change when a perturbation that allows for breaking of the
symmetry~(\ref{eq:Qsymm}), or explicitly breaks~(\ref{eq:Ssymm}), is
introduced. An exhaustive study of such perturbations is beyond the
scope of this paper. Nonetheless we wish to show that the new type of
dynamics appear at the exchange of stability when a perturbation is
introduced. This is because to a large extent our motivation has been
to show that the exchange of stability is in a sense a limit of
bifurcations that occur in realistic physical systems that reduce to
the overdamped pendulum. The perturbation is then to be understood as
the terms removed from the original nonlinear system describing the
real physical system to obtain the overdamped pendulum. Therefore we
wish to show that symmetry breaking, or other type of dynamics appear
exactly at the exchange of stability.

We use trial function of the form given in Eq.~(\ref{eq:slowtrial}) to
probe the response of the system to small additional terms. We
introduce a perturbed system
\begin{equation}
  \dot \theta + G(t) \sin \theta = F(t) + \eps H(\theta, t),
  \label{eq:genp}
\end{equation}
where $0 < \eps \ll 1$ and $H(-\theta + 2k \pi, t + T/2) = -H(\theta,
t)$ does not necessarily hold. Substituting the trial function of
Eq.~(\ref{eq:slowtrial}) into Eq.~(\ref{eq:genp}) we obtain the
equation
\begin{eqnarray}
  \dot \psi &=& -2\BZ \sin \psi  
  - \eps \Xi(\psi,t)^T \Xi(\psi, t)
  H[\theta(\psi, t), t],
  \label{eq:phipert}
\end{eqnarray}
where $\Xi(\psi, t) = \cos (\psi/2) \Pt{1}(t) + \sin (\psi/2)
\Pt{2}(t)$.  We have fixed the normalization of the Floquet solutions
so that $\Fqi{1}{1}(0)\Fqi{2}{2}(0) - \Fqi{1}{2}(0)\Fqi{2}{1}(0) =
1$\cite{isohatala09_footnote2}.  Unlike in the case of
Eq.~(\ref{eq:slowtrial}) further approximations are needed. Smallness
of $\eps$ and $\BZ$ near a bifurcation can be used to simplify the
above equation. 

\begin{figure}
  \includegraphics{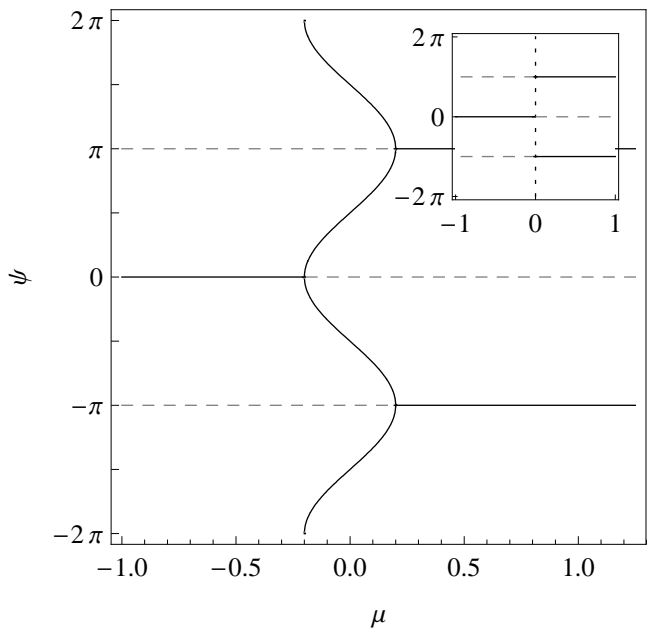}
  \caption{\label{fig:bs} Bifurcation diagram showing the equilibria
    of Eq.~(\ref{eq:psirat}) for $\eps a_2 = -0.1$, $a_k = 0$ for $k
    \neq 2$. Horizontal axis is the bifurcation parameter, $\mu =
    2\BZ$. Stable (unstable) equilibria are plotted with solid
    (dashed) lines. Inset shows the degenerate case of $a_k = 0$ for
    all $k$. Inset axis are the same as in the main figure, and the
    dotted line at $\mu = 0$ shows the marginally stable
    equilibria. The main figure demonstrates that the degenerate PB is
    replaced by a pair of supercritical PBs at $\mu = \pm 0.2$.
    Between the bifurcations the equilibria $0, \pi$ are unstable,
    while a new equilibrium with $\psi \neq 0, \pi$ that corresponds
    to symmetry broken solution of Eqs.~(\ref{eq:psirat})
    and~(\ref{eq:genp}).}
\end{figure}

First, we prove our earlier conjecture that the exchange of stability
is the $\eps \to 0$ limit of pitchfork bifurcations appearing in more
realistic systems following Eq.~(\ref{eq:Ssymm}) or equivalent.  We
consider $H = H(\theta)$ that follows the symmetry $H(-\theta + 2k\pi)
= -H(\theta)$, but contains even harmonics of $\theta$. We note that
the trial superposition $\Xi$ has the property $\Xi(-\psi + 2 k \pi, t
+ T/2) = (-1)^k \Meb \Xi(\psi, t)$. In this case, it then follows that
also Eq.~(\ref{eq:phipert}) has symmetry~(\ref{eq:Ssymm}). Using the
averaging method\cite{verhulst05}, we find that the averaged equation
for $\psi$ will have the form
\begin{equation}
  \dot \psi \simeq -2\BZ \sin \psi + \eps a_1 \sin \psi + \eps a_2
  \sin 2\psi + \cdots
  \label{eq:psirat}
\end{equation}
where $a_k$ are constants which in general are nonzero. Cosine
harmonics of $\psi$ in the averaged equation are forbidden by
symmetry. Possible bifurcations of the perturbed system can then be
qualitatively sketched by fixing $a_k$ and plotting the roots of $\dot
\psi = 0$.  As an example, if $a_2 \neq 0$ and $a_k = 0$ for $k \geq
3$, we find that the degeneracy of the pitchfork bifurcation has been
lifted.  A representative bifurcation diagram is shown in
Fig.~\ref{fig:bs}, where the equilibria of Eq.~(\ref{eq:psirat}) are
plotted for $\eps a_2 = -0.1$, $a_k = 0$ for $k \neq 2$. For
comparison, the inset shows the degenerate limit of $a_2 = 0$. It can
be seen that the degenerate PB is replaced by a pair of non-degenerate
PBs that are supercritical for $a_2 < 0$ and subcritical for $a_2 >
0$. Between the bifurcations, an equilibrium $\psi$ exists that is not
equal to $0$ or $\pi$, and thus is a symmetry broken solution of
Eq.~(\ref{eq:psirat}). In this case, the $t \to \infty$ solution
$\theta$ of the pendulum equation will be described by a superposition
of two Floquet solutions, and therefore, it too will not in general
have symmetry~(\ref{eq:Qsymm}). Thus we see that the exchange of
stability is the limit of a symmetry breaking PB for symmetrically
perturbed systems.

If the perturbation $H$ has only explicit, $T_2$-antiperiodic time
dependence, $H = H(t)$ and $H(t + T_2 / 2) = -H(t)$, then the system
will respond strongly when the perturbation frequency $\omega_2 =
2\pi/T_2$ is close to an even multiple of $\omega = 2\pi/T$. This can
be seen by noting that $\Xi^T\Xi$ has $T/2$-periodic zeroth and first
cosine harmonic of $\psi$. Therefore, the right-hand side of
Eq.~(\ref{eq:phipert}) will contain cosine harmonics of $\psi$ whose
coefficients oscillate at a frequency $|2\omega - \omega_2| \ll
1$. Using again the averaging method, these terms will not vanish but
contribute to slow, large amplitude oscillations of $\psi$. We will
later show that this effect has interesting consequences in the
problem of weak signal amplification in Josephson junctions. Note
also that this is in effect a dual of the pitchfork bifurcation
described above -- difference is that here the response follows from a
near even harmonic of the drive, not the angle $\theta$.

Finally, if the perturbation does not follow
the symmetry~(\ref{eq:Ssymm}) but still depends on $\theta$, one
expects to see the cosine terms appearing in Eq.~(\ref{eq:psirat}).
Depending on the perturbation there are several possible outcomes.
The various bifurcation scenarios can then be enumerated by selecting
the coefficients on the right-hand side of Eq.~(\ref{eq:psirat}). As
an example, the pitchfork bifurcation may become an imperfect
PB\cite{kuznetsov98}, or essentially a saddle-node bifurcation, or the
equilibria may be destroyed altogether near exchange of stability.

In summary, we described the scenario for the development of only
instability occuring in the system of type Eq.~(\ref{eq:gen})
following symmetry Eq.~(\ref{eq:Ssymm}).  The scenario follows the one
found in our previous work\cite{isohatala05:sbpend}. At a critical
point a symmetric solution of Eq.~(\ref{eq:gen}) loses stability and
another one gains it.  The exchange of stability also marks the point
where the average $\theta$ shifts from a minimum of the potential
$\cos \theta$ to a maximum or vice-versa. Exactly at the critical
point neither of the Floquet solutions is diverging (and both indeed
are still linearly independent), and hence the system will remain in
some initial superposition indefinitely. In this special case, the
symmetry~(\ref{eq:Qsymm}) need not be satisfied. This state is,
however, only marginally stable, since a perturbation will neither
decay nor diverge, and occurs only in a null set of parameter values.

We conclude by reviewing the some of the analytic
methods of approaching the overdamped pendulum equation to the exact
linearization used here. The point of commonality to these methods is
that they in one way or another consider the nonlinearity as a
perturbation. For instance, the often used technique of using single
harmonic trial function has requires the assumption that the sine term
only has the effect of changing the amplitude and phase of the
otherwise sinusoidal solution. Although the number of harmonics
included in the truncation can be increased, the equations quickly get
intractable. Averaging method can also be employed, as it indeed was
in our analysis of Eq.~(\ref{eq:psirat}), however, again the potential
term needs to be small compared to the time scale at which $\theta$
varies. In contrast, the mapping to the linear equation,
Eq.~(\ref{eq:lin}) fully retains the nonlinearity whilst still
allowing the use of tools for periodically forced linear equations,
such as Floquet theory.

Treating the nonlinearity nonperturbatively also allows us to also
approach the limit of low frequency driving. In the following section,
we derive approximate analytical formulas for the solutions of the
overdamped pendulum. Instead of studying the actual time dependence of
$\theta$, we continue with the focus on finding the critical points
where the exchange of stability occurs.

\section{\label{sec:approx}Asymptotic solution of Eq.~(\ref{eq:lin})}
We consider next approximate solutions of Eq.~(\ref{eq:gen}) in the
non-trivial limit of large $F$ and $G$, or equivalently,
low-frequency.  We introduce a large parameter $\lambda$ into the
problem by making the change $F(t) \to \lambda F(t)$, $G(t) \to
\lambda G(t)$. In the leading order of $\lambda$, we find that the
equation we need to solve comes out as
\begin{equation}
  \ddot y + \frac{\lambda^2}{4}\left(F(t)^2 - G(t)^2\right) y = 0,
  \label{eq:lin2a}
\end{equation}
where $y = q_{1}$ or $y = q_2$. Eq.~(\ref{eq:lin2a}) is found by
taking the derivative of Eq.~(\ref{eq:lin}). Keeping only terms of
order $\lambda^2$, one finds the equation~(\ref{eq:lin2a}) for both
$q_1$ and $q_2$ separately.  We note that Eq.~(\ref{eq:lin2a}) does
not share the symmetry of Eq.~(\ref{eq:lin}), however, results of the
previous section allow us to construct an approximate solution that
has the expected properties. This follows from the fact that we need
only to solve for the first half of a drive cycle $t = 0\ldots T/2$
and if need be, use Eq.~(\ref{eq:Ushift}) to obtain the complete
solution.

Standard methods of asymptotic analysis\cite{murray84:asympanal} can
be applied to Eq.~(\ref{eq:lin2a}) to find its piecewise solution in
the form (see Appendix~\ref{app:asympt}).
\begin{equation}
  y_i(t) = \frac{1}{|R(t)|^{1/4}}
  \left(
  a_i \E^{w_i\xi_i(t)} + b_i \E^{-w_i\xi_i(t)}
  \right),
  \label{eq:wkb-short}
\end{equation}
where $R(t) = (F(t)^2 - G(t)^2)/4$, $\xi_i(t) = \int_{t_i}^{t}
|R(t)|^{1/2}\;\D t$ and $w_i = 1$ ($w_i = \sqrt{-1}$) for $R(t)<0$
($R(t)>0$). We denote by $t_i$ the turning-points (points such that
$R(t_i)=0$) and by $N$ their number, \ie\ $i = 1, \ldots,
N$. Additionally, $t_0 = 0$ and $t_{N+1} = T/2$.  Coefficients $a_i,
b_i$ are solved from the initial conditions, and standard connection
formulas for adjoining subintervals are applied.  Using
Eq.~(\ref{eq:wkb-short}) it is straightforward to construct a solution
to any particular $F, G$.

Although piecewise solutions naturally can be cumbersome, we next show
that tractable formulae can be obtained for quantities of interest.
Naturally, we apply Eq.~(\ref{eq:wkb-short}) to calculating
$\tr\Mred$, as its zeros define the critical curves of the system.
For simplicity, we limit the discussion to the case of two turning
points, again, generalizations are straightforward.  With this
restriction, the trace of $\Mred$ becomes
\begin{eqnarray}
  \tr \Mred 
  &\simeq&
  - 2 \sign G(0) \sinh \left((\kappa_0 + \kappa_1) \lambda + \ln 2 \right) \cos \omega_1 \lambda \notag \\
  &-& 2 \sign G(0) \cosh \left((\kappa_0 - \kappa_1) \lambda \right) \sin \omega_1 \lambda,
  \label{eq:trm2}
\end{eqnarray}
where $\kappa_{0} = \xi_{0}(t_{1})$, $\omega_{1} = \xi_{1}(t_{2})$,
and $\kappa_1 = \xi_{2}(T/2)$.  This equation is one of the central
results of this paper, as it allows for calculating the critical
curves of overdamped pendulum equations in the non-trivial limit of
large $G$ and $F$.

In order to keep the treatment more concrete, we now fix $G = 1$ and
$F = f \sin \omega t$. We consider this restriction reasonable, since
it was demonstrated in the previous section that differential
equations of the form~(\ref{eq:gen}) exhibit the same structure as
long as $F, G$ have the appropriate symmetry. Thus, we can choose any
such forcing as a representative of the class of equations we are
considering. Further, this choice has particular relevance to the
physical systems we have in mind.

Interestingly, with this choice, Eq.~(\ref{eq:lin2a}) becomes the
Mathieu equation. In addition to constructions~(\ref{eq:wkb-short})
and~\ref{eq:trm2}), we have the known solutions at our disposal.  The
trace of $\Mred$ comes out as
\begin{equation}
  \tr \Mred \simeq -\frac{1}{\omega} S \left(\frac{f^2 - 2}{8 \omega^2}, \frac{f^2}{16 \omega^2}; \pi \right).
  \label{eq:trmm}
\end{equation}
Here $S = S(a,q;t)$ denotes the odd solution to the canonical form of
the Mathieu equation\cite{abramowitz}, $\ddot y + (a - 2 q \cos 2 t) y
= 0$, with the (non-standard) initial condition $\dot S(0) = 1$.  The
quantity $\tr \Mred$ vanishes at parameters $(a, q)$ for which $S(t)$
is periodic, whether that period was $\pi$ or $2\pi$. Values of $a$
corresponding to a periodic $S(t)$ are the Mathieu characteristic
values $b_k(q)$, $k = 1, 2, \ldots$, and thus the critical curves are
described by the equation
\begin{equation}
  \frac{f^2 - 2}{8\omega^2} = b_k\left(\frac{f^2}{16\omega^2}\right), 
  \quad k = 1, 2, \ldots
  \label{eq:mchar}
\end{equation}

In addition to Eq.~(\ref{eq:mchar}), which can be used to calculate
the parameters for which $\tr \Mred $ vanishes, for practical
applications we also need a way of computing $\tr \Mred$ for any given
$(f, \omega)$. Eq.~(\ref{eq:trmm}) is not well-suited for this purpose
since it requires the use of Mathieu functions with arbitrary
parameters. Eq.~(\ref{eq:trm2}) on the other hand has a more tractable
form as it only requires the use of elliptic integrals:
\begin{eqnarray}
  \tr \Mred 
  &=& -2 \sinh \left(\frac{\Re\{E(f^2)\}}{\omega} + \ln 2\right) \cos \frac{\Im \{E(f^2)\}}{\omega} \notag \\
  &-& 2 \sin \frac{\Im \{E(f^2)\}}{\omega},
  \label{eq:trmreda}
\end{eqnarray}
where $E$ is the complete elliptic integral of the second
kind\cite{abramowitz}.

\begin{figure}
  \begin{centering}
    \includegraphics{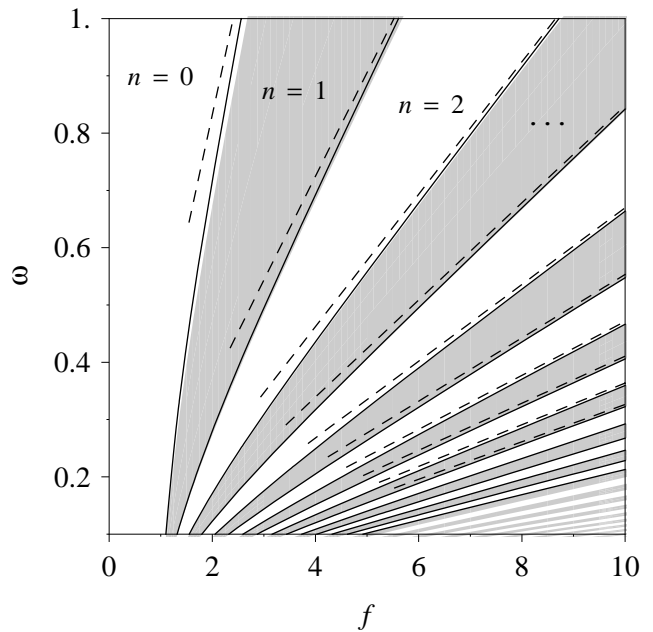}    
    \caption{\label{fig:charw} Critical curves as functions of the
      parameters $f, \omega$.  Shaded regions indicate numerically
      found solutions of Eq.~(\ref{eq:lin}) [$G(t) = 1, F(t) = f \sin
        \omega t$] for which $\langle \theta \rangle = \pi$.  Critical
      curves given by Eq.~(\ref{eq:mchar}) are plotted with solid
      lines for $k = 1 \ldots 15$. Dashed lines define the ten first
      Bessel roots [$J_0(f/\omega) = 0$], which are for clarity shown
      only for $\omega > f^{-1}$.}
  \end{centering}
\end{figure}

In Fig.~\ref{fig:charw} we have plotted the critical curves as given
by Eq.~(\ref{eq:mchar}) together with the correct numerically obtained
ones. For comparison, we have included a high-frequency approximation
to the critical lines, $J_0(f/\omega) = 0$ where $J_0$ is the Bessel
$J$ function of order zero\cite{isohatala05:sbpend}. The lines where
$f/\omega$ equals a root of $J_0$ are only in modest agreement for low
frequencies, but improves as $\omega$ is increased.
 Our new result, Eq.~(\ref{eq:mchar}) is, on the
other hand, in very good agreement for low frequencies, and is
accurate also in the opposite case of $\omega \gg 1$, especially for
large $f$. Although not plotted, the condition $\tr\Mred=0$ with
$\Mred$ given by Eq.~(\ref{eq:trmreda}) also provides very good
agreement to the computed critical curves.

A quantity that will often be needed is the average of $G \cos
\theta$, where $\theta$ is the stable solution. Using
Eqs.~(\ref{eq:avgcos}, \ref{eq:bsymm}) we find the following equation
that allows us to express $\langle G \cos \theta \rangle$ simply in
terms of $\tr \Mred$ as
\begin{equation}
  \langle G \cos \theta \rangle = 2|\BZ| = \frac{2}{T} |\arcsinh(\tr \Mred / 2)|.
  \label{avarage_cos}
\end{equation}

In summary, the main result of this section is Eq.~(\ref{eq:trm2})
whose roots give the critical curves for $F, G$ that are either large
or depend slowly on time, obey symmetry~(\ref{eq:Qsymm}), and have
exactly two distinct points $t_1, t_2$ such that $F(t_{1,2}) = \pm
G(t_{1,2})$. For the case of constant $G$ and sinusoidal $F$ our three
main results are (i) that interestingly the nonlinear equation reduces
to the Mathieu equation.  (ii) the Mathieu limit in turn allows us to
write the critical curves of the system using the Mathieu
characteristic values, with excellent agreement with the numerical
data.  (iii) Eq.~(\ref{eq:trmreda}) enables us to compute $\tr \Mred$,
and consequently also $\avgcos$ [Eq.~(\ref{avarage_cos})], in the
limit of slow external drive. This last result will be used in the
following section.

\section{\label{sec:applications}Applications to physical systems}
In this section we touch upon three physical problems to which the
theory developed above can be directly applied: Rectification of
microwaves in lateral semiconductor
superlattices\cite{dodin98b,alekseev05:lsslcr}, amplification of
high-frequency signals in Josephson point contacts, and modeling of
Josephson junctions with critical current modulation\cite{chesca08}.

\subsection{\label{sec:lssl}Semiconductor superlattices}
\begin{figure}
  \includegraphics{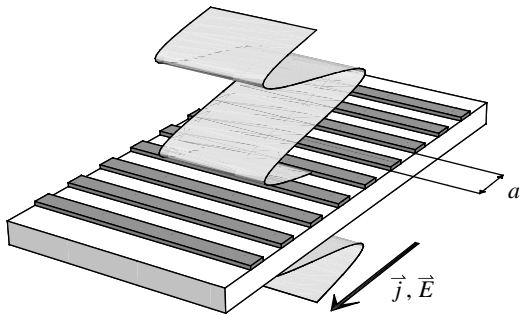}
  \caption{\label{fig:schem} Schematic figure showing the ac-driven
    lateral superlattice.  Electromagnetic wave is incident normal to
    the plane of 2D gas of conduction electrons. Electric field of the
    wave $\vec{E}(t)$ induces the current density $\vec{j}(t)$ flowing
    perpendicular to the superlattice layers. Along the direction of
    the current electrons experience a periodic potential with period
    $a$.}
\end{figure}
 We present the problem in the form it was
introduced in Ref.~\onlinecite{alekseev05:lsslcr}.  A schematic figure
showing the geometry of the problem is given in
Fig.~\ref{fig:schem}. Plane electromagnetic wave incident on a lateral
superlattice is considered.  Electric field is polarized along the
superlattice axis, that is, parallel to the direction of the
current. The electron transport is studied by considering a single
miniband with the standard tight-binding energy-quasimomentum
dispersion relation\cite{wacker02:ssreview}.  The electron
distribution follows the Boltzmann transport equation. From there, one
is able to write ordinary differential equations for ensemble averaged
electron velocity and energy. These form the well-known superlattice
balance equations\cite{ignatov76, wacker02:ssreview}. We present the
equations here in their scaled form, in which the the maximum
(minimum) electron velocity $v$ and energy $w$ correspond to the value
$+1$ ($-1$).  Electric field inside the superlattice, denoted $u$, is
also appropriately scaled.
\begin{subequations}
  \begin{eqnarray}
    \dot v &=& - u w - \Gamma v, \label{eq:belssl_v} \\
    \dot w &=& \phantom{-} u v - \Gamma (w - \weq). \label{eq:belssl_w}
  \end{eqnarray}
  \label{eq:belssl}
\end{subequations}
First equation of the set describes the balance between electron
acceleration by the electric field and deceleration due to scattering,
while the second describes the electron energy gain and dissipation
due to scattering processes.  The current density $j$ is related to
the average velocity $v$ by $j \propto e N_s v$, where $e$ is the
elementary charge and $N_s$ is the areal density of 2D electron gas.
The nonlinearity is controlled by the parameter $\Gamma$: $\Gamma
\propto (\gamma / N_s)^{1/2}$, where $\gamma$ is rate of electron
scattering -- a high density of the electron gas corresponds to a
large nonlinearity, or small $\Gamma$.  Constant $\weq$ is the scaled
equilibrium energy, whose value we set to $-1$ for convenience.

The interaction of the incident electromagnetic radiation with the
conduction electrons is taken into account by employing the Maxwell
equations with appropriate boundary conditions.  Approximating the
lateral superlattice as an infinite conduction sheet, the scaled
electric field entering the equations, $u$, becomes\cite{dodin98b}
\begin{equation}
  u = -u_0 \cos \Omega t - \Gamma^{-1} v,
  \label{eq:ulssl}
\end{equation}
where $u_0$ and $\Omega$ are the amplitude and frequency of the
external electric field. With the above, Eqs.~(\ref{eq:belssl}) are
rendered nonlinear.  Unlike in the case of bulk superlattices where
the relation between the total electric field $u$ and the average
velocity $v$ is an additional differential
equation\cite{alekseev96:dissp-chaos-ssl, alekseev98:spont-dc}, here
the equation is algebraic.

The overdamped pendulum equation is obtained via a formal change of
variables, $v = -A \sin \theta$, $w = -A \cos \theta$. In the
physically interesting limit of $\Gamma \ll 1$, the dimensionality of
the system can be reduced (see Appendix~\ref{sec:pendlim}).  The
following equation for $\theta$ is obtained:
\begin{equation}
  \dot \theta + \left(\frac{\avgcos}{\Gamma} 
  + \frac{\Gamma}{\avgcos}\right) \sin \theta = u_0 \cos \Omega t.
  \label{eq:latpend}
\end{equation}

One of the primary interests is the appearance
of a spontaneous dc voltage.  The term ``rectification'' was used for
the conversion of applied ac irradiation into a dc field $\langle u
\rangle \neq 0$ and current $\langle j \rangle \propto \langle v
\rangle \neq 0$, via the nonlinearity of these nanostructures. In
terms of the variables $A, \theta$, a prerequisite for such a current
to appear is that a limit-cycle does not follow Eq.~(\ref{eq:Qsymm}),
\ie\ $\theta(t + T/2) \neq -\theta(t) + 2 k \pi$. However, since the
governing equations are in fact symmetric in the sense of
Eq.~(\ref{eq:Qsymm}), and based on the findings of
Sec.~\ref{sec:symbr}, the breaking of symmetry implies a pitchfork
bifurcation. Consequently, rectification is expected in the real
physical system when parameters are such that they correspond to the
exchange of stability in the overdamped pendulum.

In the earlier work\cite{alekseev05:lsslcr}, we have considered a
simplified pendulum equations in which the contribution of $\avgcos$
was ignored, that is the overdamped pendulum Eq.~(\ref{eq:gen}) with
$G=1$ and $F(t)=u_0 \cos\omega t$.  Our analytic analysis of that
equation in \onlinecite{alekseev05:lsslcr} was limited to the high
driving frequency limit where the instability occurs in the vicinity
of $J_0(u_0/\omega) = 0$ (\cf\  Fig.~\ref{fig:charw}).
Comparing with results of numerical solutions of the superlattice
balance equations, we observed that the rectification indeed exists
nearby the Bessel roots. Here we apply the theory developed in the
previous sections to Eq.~(\ref{eq:latpend}) in order to find the
regions of instability in a wider parameter space, including the case
of low driving frequencies.

We can solve the functional-differential equation~(\ref{eq:latpend})
by considering the equation $\dot \theta + K \sin \theta = u_0 \cos
\Omega t$. We require that $K$ equals the coefficient of sine in
Eq.~(\ref{eq:latpend}):
\begin{equation}
  K = \frac{\avgcos(K)}{\Gamma} + \frac{\Gamma}{\avgcos(K)}.
  \label{eq:AB}
\end{equation}
Eq.~(\ref{avarage_cos}) allows us to write $\avgcos$ in terms of $\tr
\Mred$, and further, Eq.~(\ref{eq:trmreda}) gives $\tr \Mred$ using
well-known special functions. Thus, roots of Eq.~(\ref{eq:AB}) can be
easily computed numerically. 

Alternatively, parameter space structure of Eq.~(\ref{eq:latpend}) can
be studied by the following way. We introduce a simple change of
variables $(f, \omega) \to (u_0, \Omega)$: $(u_0, \Omega) = (A\Gamma^{-1}
+ A^{-1}\Gamma)\cdot(f, \omega)$, where $A=\avgcos(f,\omega) =
2|\BZ(f, \omega)|$ can be found following Eq.~(\ref{avarage_cos}).
Using this transformation, parameter space structure of
Eq.~(\ref{eq:latpend}) in variables $(u_0,\Omega)$ can be studied for
any $\Gamma$ by calculating a single dataset of values $(f, \omega,
\BZ(f, \omega))$ from the pendulum equation (\ref{eq:gen}) with $G=1$
and $F(t)=f \cos\omega t$.
\begin{figure}
  \begin{centering}
    \includegraphics{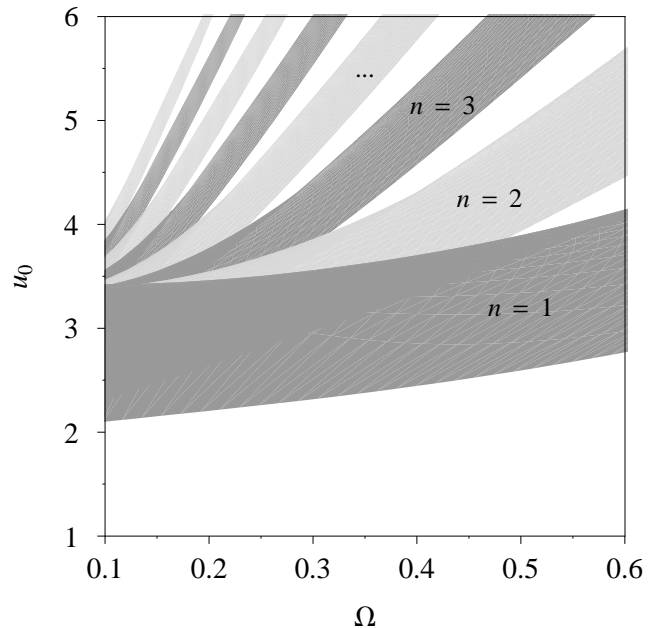}
    \caption{\label{fig:lssl} Parameter space of
      Eq.~(\ref{eq:latpend}) found for $\Gamma = 0.2$ employing the
      asymptotic solutions discussed in Sec.~\ref{sec:approx}.
      Shading is used to distinguish different branches of solutions;
      white areas correspond to no valid solution ($\avgcos < \Gamma$)
      or branch other than $1, \ldots, 8$. }
  \end{centering}
\end{figure}
For the case of $\Gamma = 0.2$ this procedure was applied to plot
Fig.~\ref{fig:lssl}, where branches of the solutions with $n = 1,
\ldots, 8$ are displayed. The pendulum limit of the superlattice
balance equations [Eq.~(\ref{eq:latpend})] becomes invalid as $\avgcos
\lesssim \Gamma$, and thus we have chosen not to plot regions
corresponding to $\avgcos < \Gamma$.  In the case of the simple
overdamped pendulum, the position of, say, the $n$th parameter space
region of solutions is determined solely by the external drive
amplitude and frequency [Fig.~\ref{fig:charw}].  In
Eq.~(\ref{eq:latpend}) the average cosine affects values of $(u_0,
\Omega)$ that admit a solution corresponding to a particular $n$.
Account of the $\avgcos$ dependence has the effect of shifting the
parameter space points corresponding to high values of $\avgcos$ to
the higher values of $u$, which can be seen from the change of
variables $(f, \omega) \to (u_0, \Omega)$. For sufficiently low
$\Gamma$, this shift is enough to make several regions with different
$n$ co-existing at fixed $(u_0, \Omega)$.  The overlap of the different
regions of solutions can be seen in Fig.~\ref{fig:lssl}.

Comparing to Fig.~2 of Ref.~\onlinecite{alekseev05:lsslcr}, we find
that the branches $n = 2, 3, 4$ match the symmetry broken regions
found using direct numerical simulation of the superlattice balance
equations\cite{isohatala09_footnote1}.  Branch $n = 1$ in
Fig.~\ref{fig:lssl} falls under the branch $n = 0$ (not plotted),
which suggests that a symmetry broken region might have been missed in
the earlier work. Indeed, our subsequent numerical simulations confirm
that there is a region of symmetry broken solutions associated with
the instability between regions $n = 0$ and $n = 1$. It apparently was
not detected because the initial conditions preferred the $n=0$
solution. This immediately demonstrates the usefulness of our analytic
results.

The abstract theory developed in Sec.~\ref{sec:symm} gives additional
insight into the dynamics of semiconductor superlattice electrons. In
Sec.~\ref{sec:symm} we showed that the integer $n$ was connected to
the number of rotations of $\theta$, and consequently oscillations of
periodic functions depending on $\theta$: in short, $n$ counts the
number of full oscillations of $\sin \theta$ and $\cos \theta$ across
half the period of the drive. Since here $\sin \theta$ and $\cos
\theta$ correspond to the average electron velocity and energy,
respectively, we can now connect the dynamics of the charge carriers
in the periodic potential to the stability of the system.  At the
instability the integer $n$ changes by one, and so the instability in
fact marks the region where the state with $n + 1$ oscillations of
energy or current become favourable to the state with $n$
oscillations.  Further, each of the branches $n = 0, 1, \ldots$ are
characterized by the number of current oscillations across half of the
drive period.

 In summary, the theory developed in the
previous sections devoted to the mathematical analysis of
Eq.~(\ref{eq:gen} allowed us to analytically probe the dynamics of
Eqs.~(\ref{eq:belssl}) in the physically interesting limit of $\Gamma
\ll 1$, $\omega \ll 1$. This limit is equivalent to very high
nonlinearity, and has been largely inaccessibly analytically.
Importantly, the analytical results revealed the significant
multistability in this system.  We found that several branches of
solutions can coexist at same parameters $(u_0, \omega)$. The
analytical results also directly suggested, that a large region of
rectification was missed.  The region of rectification was found to be
much larger than expected in the case of $\Gamma = 0.2$. Our new
analytical findings are also consistent with the previous result that
$\Gamma \lesssim 0.4$ is required for observing rectification.

\newcommand{\Abs}{\mathcal{A}}
\newcommand{\AbsJJ}{{\Abs_\text{JJ}}}
\newcommand{\AbsSL}{{\Abs_\text{SL}}}

\subsection{\label{sec:jj}Josephson junctions}
As a second application, we consider the problem of a high-frequency
gain (negative absorption) in microwave irradiated Josephson point
contacts.  The corresponding motion equation for the Josephson phase
difference $\theta$ is\cite{likharev-c10-11, velichko99}
\begin{equation}
  \dot \theta + \sin \theta = f \sin \omega t + \eps \cos \Omega t.
\label{eq:jj-motion}  
\end{equation}
In addition to the driving current $f \sin \omega t$, a probe current
$I_p = \eps \cos \Omega t$ has been added. The probe amplitude $\eps$
is assumed small and the frequency $\Omega$ is incommensurate to the
drive $\omega$.

Having introduced a weak probe current $I_p$,
we wish to find the power absorbed by the junction, and its dependence
on the frequency of $I_p$. The absorbed power is described by $\Abs =
\langle U \cdot I_p \rangle$, where $U$ is the voltage across the
junction. In terms of pendulum variables the absorption takes the
form $\AbsJJ = \eps \langle \dot{\theta}(t)\cdot\sin\Omega t
\rangle$. For $\AbsJJ < 0$ we have gain, \ie\ a weak signal $\eps \sin
\Omega t$ will be amplified.  The system placed in a cavity will
radiate at frequencies for which $\AbsJJ < 0$.  Linearizing
Eq.~(\ref{eq:jj-motion}) and using the approach similar to
\onlinecite{kuzmin80, *kuzmin80-orig}, we find
\begin{equation}
  \AbsJJ = \frac{\eps^2}{2} \sum_{k = -\infty}^\infty b_{-k}d_k
  \frac{\Omega^2}{\Omega^2 + (\avgcos + 2\I k \omega)^2},
  \label{eq:absjj}
\end{equation}
where
\begin{equation}
  \label{eq:coefs}
  \E^{\pm \int_0^t\left[\cos \theta(t') - \avgcos\right]\; \D t'}
  =
  \sum_{k = -\infty}^\infty \E^{2\I k \omega t} 
  \left\{
  \begin{array}{l}
    b_k \\
    d_k 
  \end{array}
  \right.
  . 
\end{equation}
Derivation of Eq.~(\ref{eq:absjj}) is presented in
Appendix~\ref{app:absjj}.

Near exchange of stability, $\avgcos = 2 |\BZ| \to 0$, the expression
for absorption consists of terms of the form $1/(\Omega^2 - 4
k^2\omega^2)$.  These diverge when $\Omega \to 2k \omega$ ($k$ is an
integer), and thus we expect to see strong gain for probe frequencies
near an even multiple of the pump frequency when $\BZ \sim 0$.

\begin{figure}
  \begin{centering}
    \includegraphics{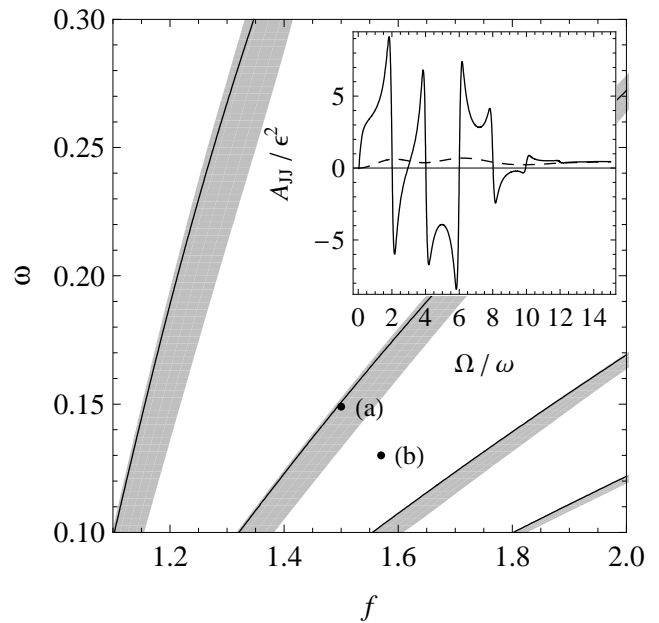}
    \caption{\label{fig:gain} Regions of negative absorption computed
      from Eq.~(\ref{eq:absjj}). Shaded regions correspond to negative
      absorption for some probe frequency in the range $\Omega =
      0\ldots 11.5 \cdot \omega$. Solid lines indicate exchange of
      stability as given by Eq.~(\ref{eq:trmreda}).  Points labeled
      (a) and (b) indicate parameters for the absorption profiles in
      the inset. Inset: Typical absorption profiles for low pump
      frequencies. Solid and dashed curves correspond to points (a),
      $f = 1.5$, $\omega = 0.149$, (b), $f = 1.57$, $\omega = 0.13$,
      respectively.}
  \end{centering}
\end{figure}

Again, results of the previous section can be applied to find $(f,
\omega)$ that result in such strong response. The range of parameters
where gain occurs is rather narrow, and hence an analytic first
approximation is useful. In Fig.~\ref{fig:gain} we plot the regions of
negative absorption using Eq.~(\ref{eq:absjj}) together with the
critical curves as given by Eq.~(\ref{eq:trmreda}).  Coefficients
$b_k$, $d_k$ were computed numerically from the Floquet solutions
using the fact that the expansions in Eq.~(\ref{eq:coefs}) are equal
to $(a {\Pt{}}^T\Pt{})^{\pm 1}$ [\cf\ App.~\ref{app:absjj}], where
$\Pt{}$ is oscillating part of the unstable Floquet solution and $a =
(\Pt{}(0)^{T} \Pt{}(0))^{-1}$.  We also made several runs computing
$\AbsJJ$ directly from Eq.~(\ref{eq:jj-motion}).  While this direct
procedure is in general more slow, its results are in a good agreement
with the results obtained employing Eqs.~(\ref{eq:absjj})
and~(\ref{eq:coefs}).

To illustrate the strong dependence of the absorption on the
parameters we have also plotted typical absorption profiles,
dependence of $\AbsJJ$ on $\Omega$, in the inset of
Fig.~\ref{fig:gain}: the profile forms strong peaks just as the
critical curve for exchange of stability is approached.  Gain peaks
are located in the vicinities of even harmonics $2k \omega$ of the
driving current.  Therefore the system can radiate at frequencies
close to $2k \omega$.  This effect is very different from ordinary
generation of harmonics. Really, following
symmetry~(\ref{eq:Qsymm}) overdamped pendulum can generate only odd
harmonics $(2k+1) \omega$ of the strong pump while even harmonics are
forbidden.  From physical point of view, generation at (exactly) odd
harmonics is a spontaneous process, whereas radiation at frequencies
nearby even harmonics is stimulated emission.  From the viewpoint of
bifurcations, the effect of gain in overdamped pendulum represents a
reminiscent of symmetry breaking bifurcation in strongly damped second
order pendulum Eq.~(\ref{eq:ddp}), where the bifurcation results in
appearance of even harmonics.  Despite here even harmonics are
forbidden by symmetry, nontrivial amplification of additional weak
signal does exist at pump amplitudes and frequencies close to those
necessary to realize real symmetry breaking.

Interestingly, similar gain profiles, centered near a characteristic
Bloch or cyclotron frequency and its harmonics, attract much attention
in physics of semiconductor superlattices, where them termed
dispersive gain profiles \cite{sekine05, hyart09_magnetic}.  Moreover,
the dispersive gain profiles were found in the models of ac-driven
semiconductor superlattices as well\cite{hyart08,hyart09_quasistatic}.

Recently, relatively strong coherent
electromagnetic radiation of technologically important terahertz
frequency band has been observed from dc-driven high-$T_c$
superconductors\cite{minami09} and arrays of niobium point
contacts\cite{song09}. Physical mechanisms responsible for the
observed radiation are still subject of intensive
debates. Nevertheless, we notice that our simple model of a pure
ac-driven junction gives a natural framework to generalization of
dc-drive configurations considered experimentally, and in this respect
a further development of this theory may have a beneficial applied
aspect.

Finally, as a separate remark, we note that the general form of
Eq.~(\ref{eq:gen}) with $G(t) \neq \text{constant}$ and the associated
symmetry do have relevance to more complex models of Josephson
junctions.  Recently, a pendulum equation was used to model observed
current-voltage characteristics of a type of Josephson junction with a
strong applied microwave field\cite{chesca08}. The magnetic field
induced changes to the critical current were considered significant
enough to be incorporated into the model. The resulting equation has,
in the low-frequency drive case, a form corresponding to
Eq.~(\ref{eq:gen}) and, in the absence of a dc current component, also
the symmetry considered in current paper.

\section{Conclusions}
We have studied the periodically driven overdamped
equation~(\ref{eq:gen}) with periodic coefficients by using a type of
Pr\"{u}fer transformation. The linear form Eq.~(\ref{eq:lin}) allowed
for easy analysis of the nonlinear system.  We showed that if the
system is driven by external forcing of the form Eq.~(\ref{eq:Ssymm})
there exists only one type of instability in the system.  The
instability was identified as an exchange of stability between two
periodic solutions, and it was found to be essentially a degenerate
form of a pitchfork bifurcation. We showed that the degeneracy of the
bifurcation is lifted when additional terms that also follow the
symmetry are added, thus proving that the exchange of stability is in
a way a precursor to symmetry breaking bifurcations in realistic
physical systems.  Also, we found that near the exchange of stability,
the overdamped pendulum responds strongly to perturbations whose
frequency is close to an even multiple of the drive frequency.  We
further used the linear form to find explicit solutions to the
problem, and used them to construct a condition for the appearance of
the instability.  For the simple choice of $F(t) = f \sin \omega t$,
$G(t) = 1$ the instabilities of Eq.~(\ref{eq:gen}) can be well
described by the Mathieu equation. We wish to emphasize, that up to
the knowledge of the authors, no such low-frequency solutions have
been constructed so far.

From our purely mathematical findings there is then a natural
cross-over into the field of physics. As was shown, near exchange of
stability even weak perturbations can generate symmetry breaking
bifurcations, or other types of strong response. Thus, regions close
to this bifurcation in real physical systems are expected to show
novel phenomena.  We applied our methods directly the real physical
problems, namely rectification of pure ac irradiation in lateral
semiconductor superlattices and amplification of a weak signal in
Josephson point contacts.

For lateral semiconductor superlattices, making use of our analytical
methods, we were able to construct the parameter and phase space
structure of this system. Excellent agreement was found with our
previous numerical results\cite{alekseev05:lsslcr}, with the
additional finding that the system exhibits strong multistability:
parameter space consists of overlapping regions each carrying a
solution that is characterized by the extent of oscillations of the
electron gas average energy and velocity. Each of these regions was
found to have an associated region of instability which in the low
frequency drive and strong nonlinearity limit corresponds to
rectification of the incident ac electric field -- that is, the effect
where the nonlinear interaction between the 2D electron gas and the
external ac field generates a directed current and voltage across the
superlattice. The analytical results revealed the complex way in which
the multistability and the related regions of instability appear, and
suggested that large regions of rectification were missed in earlier
simulations. In fact, rectification persists for higher frequencies
than was previously thought.  The new results supported the earlier
finding that fairly high electron mobilities are required for
rectification to appear.


Second physical system that we considered was amplification of a weak
signal in Josephson point contacts. We found (i) a direct and striking
correspondence between gain, that is, negative absorption and the
exchange of stability bifurcation -- strong gain was shown to appear
in a narrow region just around the point where loss of stability
occurs, when also the weak signal frequency was close to an even
multiple of the strong ac driving current. Secondly, (ii) our
analytical results make it possible to find the regions of gain very
accurately, or allow for computing them with little cost. Further,
(iii) our findings on the dynamics of overdamped pendulum suggest
insight into the physical process of amplification -- the overdamped
pendulum exhibits only odd harmonics which follows from the
constraints set by the symmetry, yet, we found that the system is
expected to radiate when the weak signal to be amplified is tuned to
even harmonics.

 Finally, we would like to point out
intriguing similarities between gain profiles found here in the model
of ac-driven Josephson junction (ac-driven overdamped pendulum) and
dispersive gain profiles recently described in the models of ac-driven
bulk [not lateral] semiconductor superlattices in \cite{hyart08,
  hyart09_quasistatic}. In these works\cite{hyart08,
  hyart09_quasistatic}, a superlattice is in essence modeled by two
standard balance equations for electron velocity and miniband
energy\cite{ignatov76},
that is, Eqs.~(\ref{eq:belssl}) without $v$ dependence in $u$.
Taking into account some earlier findings\cite{alekseev02a}, we
speculate that bifurcations of overdamped pendulum also determine
conditions for high-frequency gain in ac-driven bulk superlattices,
described by the standard balance equations, in the limits of high
frequencies and rare collisions.  A detailed comparison of properties
of amplification in ac-driven Josephson junctions with the properties
of dispersive gain in ac-driven semiconductor superlattices
\cite{hyart08, hyart09_quasistatic} in the limit of weak dissipation,
however, goes beyond the scope of the present paper.

\begin{acknowledgments} We are thankful to Alex Zharov and Andrei Malkin for
discussions on nonlinear dynamics in lateral superlattices, Boris
Cheska and Marat Gaifullin on experimental aspects of pendulum-like
dynamics in Josephson junctions, Leonid Kuzmin, Alexander Klushin, and
Kazuo Kadowake -- on amplification and generation of high-frequency
radiation in Josephson junctions and their arrays, Timo Hyart -- on
correspondence between semiconductor superlattices and Josephson
junctions, and Sasha Balanov for bifurcations in pendulum.

We thank Feo Kusmartsev and Erkki Thuneberg for a constant
encouragement of this activity. This research was partially supported
by AQDJJ Programme of European Science Foundation.
\end{acknowledgments}

\appendix
\section{\label{app:asympt}Construction of the asymptotic solution}
Here we briefly outline how the approximate asymptotic solution of
Eq.~(\ref{eq:lin}) is constructed using asymptotic solutions of
Eq.~(\ref{eq:lin2a}). Without loss of generality we can assume that
$R(t = 0) < 0$. The interval $t = 0 \ldots T/2$ is divided into
subintervals $I_j = (t_j, t_{j + 1})$ where $t_j$, $j = 1 \ldots N$ is
a turning point and $N$ is the total number of turning points. We
additionally set $t_0 = 0$ and $t_{N + 1} = T/2$. Solutions on each of
the intervals are approximately given by the usual WKB solutions
\begin{subequations}
\begin{eqnarray}
  y_{2k}(t)  &=& 
  \frac{
    A_{2k} \E^{\lambda \xi_{2k}(t)} + \frac{1}{2} B_{2k} \E^{-\lambda \xi_{2k}(t)}
  }{
    (-R(t))^{1/4}
  }, \label{eq:wkb_a} \\
  y_{2k+1}(t) &=& 
  \frac{A_{2k+1}}{R(t)^{1/4}} 
  \sin \left( \lambda \xi_{2k+1}(t) + \frac{\pi}{4} \right) \notag 
   \\
    &+&
  \frac{B_{2k+1}}{R(t)^{1/4}} 
   \cos \left( \lambda \xi_{2k+1}(t) + \frac{\pi}{4} \right), \\
  \xi_{k}(t) &=& \int_{t_k}^t \sqrt{|R(t')|} \; \D t'.
 \end{eqnarray}
\label{eq:wkb}
\end{subequations}
Each of the above solutions $y_k$ is only valid in its corresponding
interval $I_k$ -- when $R(t) < 0$, solutions have the exponential form
$y_{2k}$ and when $R(t) > 0$, the oscillatory form $y_{2k+1}$ is the
appropriate solution [\cf\ harmonic oscillator $\ddot y + r y = 0$,
  $r$ real, solutions oscillate for $r > 0$ and converge or diverge
  exponentially when $r < 0$].

Connection formulas for the coefficients $C_k = (A_k, B_k)^T$ can be
derived by solving the problem at the turning points.  For first order
zeros of $R(t)$, \ie\ roots $t^*$ such that $\dot R(t^*) \neq 0$, the
approximate solutions $x_k$ around $t = t_k$ are given in terms of the
Airy functions $\Ai, \Bi$ as
\begin{equation}
  x_k = A^*_k [\dot \phi_k]^{-1} \Ai(\lambda^{2/3} \phi_k) 
    + B^*_k [\dot \phi_k]^{-1} \Bi(\lambda^{2/3} \phi_k) 
\end{equation}
with
\begin{equation}
  \phi_k = \left(\frac{3}{2} \int_{t_k}^t [-R(t')]^{1/2}\; \D t'\right)^{2/3},
\end{equation}
where $A^*_k$, $B^*_k$ are constants.  Note that above we need to
raise a complex number $z$ to power $2/3$, where $z$ is either real or
pure imaginary.  Here, the argument of $z^{2/3}$ is chosen to be $0$
or $\pi$, \ie\ so that $\phi_k$ is real. Doing so it follows that if
$R(t)$ is increasing (decreasing) around $t_k$, then $\phi_k(t)$,
$\phi_k(t_k) = 0$, is continuous and decreasing (increasing) around
$t_k$.  Away from the turning points the functions $x_k$ asymptote
into the WKB solutions given in Eq.~(\ref{eq:wkb}).  This allows one
to write a linear relationship between $C_k$ and $C_{k+1}$: $C_{k+1} =
W_k C_k$, where
\begin{subequations}
\begin{eqnarray}
  W_{2k} &=&
  \begin{Mtx}
    2 \exp \left( \kappa_{2k} \lambda \right ) & 0 \\
    0 &  \frac{1}{2} \exp \left( -\kappa_{2k} \lambda \right )
  \end{Mtx}, \\
  W_{2k-1} &=&
  \begin{Mtx}
    \cos \omega_{2k-1} \lambda & - \sin \omega_{2k-1} \lambda \\
    \sin \omega_{2k-1} \lambda &   \cos \omega_{2k-1} \lambda 
  \end{Mtx}.
\end{eqnarray}
\label{eq:wconn}
\end{subequations}
Here, $\omega_{2k-1} = \xi_{2k-1}(t_{2k})$ and $\kappa_{2k} =
\xi_{2k}(t_{2k+1})$. 
Denoting $Y(t) = (y(t), \dot y(t))^T$, we wish to construct a matrix
$V$ so that $Y(T/2) = V \cdot Y(0)$. Above we have derived connection
formulas for the superposition coefficients $C_k$ across the whole
interval $0 \ldots T/2$, so all we need are matrices that map the
initial values of $Y$, $Y(0)$, to the coefficients $C_0$, and the last
coefficients $C_N$ to the end values $Y(T/2)$. In other words, we need
$S_0$ and $S_1$ so that $C_0 = S_0 \cdot Y(0)$ and $Y(T/2) = S_1 \cdot
C_N$.  Using Eq.~(\ref{eq:wkb_a}), $(y_0(0), \dot y_0(0))^T$ can be
written in terms of the coefficients $C_0 = (A_0, B_0)^T$. After
solving $(A_0, B_0)$ the matrix $S_0$ can be read out.  Similarly,
using the same equation and writing out $(y_N(T/2), \dot y_N(T/2))^T$,
the coefficient matrix $S_1$ is found. We obtain
\begin{subequations}
  \begin{eqnarray}
    S_0 &=& 
    \begin{Mtx}
      \frac{1}{2}r_0 - \frac{r_1}{8\lambda r_0^5} &
      \frac{1}{2\lambda r_0}
      \\
      r_0 + \frac{r_1}{4\lambda r_0^5} &
      -\frac{1}{\lambda r_0}
    \end{Mtx},
    \\
    S_1 &=&
    \begin{Mtx}
      \alpha \frac{1}{r_0} &
      \frac{1}{\alpha} \frac{1}{2 r_0}
      \\
      \alpha \left[ \lambda r_0 + \frac{r_1}{4r_0^5} \right] &
      \frac{1}{\alpha} \left[ -\frac{\lambda r_0}{2} + \frac{r_1}{8 r_0^5} \right]
    \end{Mtx},
  \end{eqnarray}
\end{subequations}
where $r_0 = |R(0)|^{1/4}$, $r_1 = \dot R(0)$, and $\alpha = \exp(\lambda
\kappa_N)$.

Using Eqs.~(\ref{eq:wkb}) and~(\ref{eq:wconn}) the solution at $t =
T/2$ can be obtained in terms of the initial values $y(0), \dot y(0)$.
\begin{equation}
  Y(T/2) = S_1 W_{N-1} W_{N-2} \cdots W_0 S_0 Y(0).
  \label{eq:Ymap}
\end{equation}
Finally, to get the value of $Q$ at $t = T/2$, Eq.~(\ref{eq:Ymap}) is
applied with the initial conditions
\begin{equation}
  Y(0) = \qvec{q_1(0)}{\dot q_1(0)}, \quad
  Y(0) = \qvec{q_2(0)}{\dot q_2(0)}
\end{equation}
to get $q_1(T/2)$ and $q_2(T/2)$, respectively.

\section{\label{sec:pendlim}Overdamped pendulum limit of lateral superlattice balance equations}
In this appendix we present the derivation of Eq.~(\ref{eq:latpend}).
We start with the superlattice balance equations,
Eqs.~(\ref{eq:belssl}) with an electric field following
Eq.~(\ref{eq:ulssl}),
\begin{subequations}
\begin{eqnarray}
  \dot v &=& -(u_\text{in}(t) - \Gamma^{-1} v) w - \Gamma v, \\
  \dot w &=& \phantom{-}(u_\text{in}(t) - \Gamma^{-1} v) v - \Gamma(w - w_\text{eq}).
\end{eqnarray}
  \label{eq:lsslbe}
\end{subequations}
Here, $v, w$ are average electron velocity and energy scaled to
dimensionless units and into range $-1 \ldots 1$, $u_\text{in}$ is the
scaled incident electric field which is taken to be of the form
$u_\text{in} = -u_0 \cos \Omega t$. For details we refer the reader to
Ref.~\onlinecite{alekseev05:lsslcr}.

The pendulum form is obtained by making the substitutions $v = -A \sin
\theta$, $w = -A \cos \theta$. From Eqs.~(\ref{eq:lsslbe}) we get
differential equations for $A, \theta$:
\begin{subequations}
  \begin{eqnarray}
    \dot A      &=& -\Gamma A - w_\text{eq} \Gamma \cos \theta, \label{eq:lsslpenda} \\
    \dot \theta &=& -\left(\frac{A}{\Gamma}-w_\text{eq} \frac{\Gamma}{A}\right)\sin \theta 
                    - u_\text{in}(t). \label{eq:lsslpendb}
  \end{eqnarray}
\end{subequations}
We are interested in the case $\Gamma \ll 1$.  It follows that, since
$\dot A \propto \Gamma$, we may consider $A$ a slow variable. Provided
that $\Gamma \lesssim \Omega$ we can say that to a first approximation
$A$ is constant over one cycle of the drive field $T$, $T =
2\pi/\Omega$.  Taking the average of Eq.~(\ref{eq:lsslpenda}) we get
$A \approx -w_\text{eq} \langle \cos \theta \rangle$. Finally, setting
$w_\text{eq} = -1$ for simplicity and substituting $A \to \langle \cos
\theta \rangle$ into Eq.~(\ref{eq:lsslpendb}) we get
Eq.~(\ref{eq:latpend}).

\section{\label{app:absjj}Derivation of formula for Josephson junction absorption}
\newcommand{\dtheta}{{\delta \theta}} 
\newcommand{\omegap}{\Omega}
\newcommand{\omegad}{\omega}
We begin by linearizing Eq.~(\ref{eq:jj-motion}) in $\eps$: Setting
$\theta \to \theta + \dtheta$, we get
\begin{equation}
  \dot \dtheta + \cos \theta \dtheta = \eps \cos \omegap t,
\end{equation}
where $\theta$ follows Eq.~(\ref{eq:jj-motion}) with $\eps = 0$.
The above has the exact solution 
\begin{equation}
  \dtheta = \eps \E^{ - \int_0^t \cos \theta (t') \;\D t'}
            \int \limits_{-\infty}^t \E^{\int_0^{t'} \cos \theta (t'') \;\D t''} \cos \omegap t' \;\D t'.
\end{equation}
We note that since the periodic solutions $\theta$ are symmetric,
$\theta(t + T/2) = -\theta(t) + 2k\pi$, then $\cos \theta$ is
$T/2$-periodic.  Thus, we rewrite the exponentials appearing in
$\dtheta$ as a Fourier series with only even harmonics of
$\omegap$. This gives Eq.~(\ref{eq:coefs}). Using these expansions,
and denoting $\avgcos = \beta$ for brevity, $\dtheta$ becomes:
\begin{eqnarray}
  \dtheta &=& \eps \sum_{k,k'} b_k \E^{\I 2 \omegad k t} \E^{-\beta t} 
              \int d_{k'} \E^{\beta t} \E^{\I 2 \omegad k' t} \cos \omegap t \; \D t \notag \\
          &=& \eps 
              \sum_{n}\sum_{k} b_{n-k} d_k \E^{\I 2 \omegad n t}  \notag \\
          &\phantom{=}& \times
              \frac{(\beta + 2 \I k \omegad) \cos \omegap t + \omegap \sin \omegap t}{\omegap^2 + (\beta + 2 \I k \omegad)^2}.
              \label{eq:dthetaser}
\end{eqnarray}
Next, we calculate $\AbsJJ$. Since $\cos \omegap t$ harmonic is not
present in $\theta$, $\AbsJJ$ becomes $\eps \langle \cos \omegap t
\cdot \dot \dtheta \rangle$. By partial integration, we get $\AbsJJ =
\eps \omegap \langle \sin \omegap t \cdot \dtheta \rangle$. The
averaging then simply picks from $\dtheta$ the coefficient of $\E^{2
  \I n \omegad t} \sin \omegap t$, $n = 0$, divided by two. Upon
inspection of Eq.~(\ref{eq:dthetaser}), Eq.~(\ref{eq:absjj}) follows.

Finally, we note that the expressions in Eq.~(\ref{eq:coefs}) are
equal to $(a\Pt{}^{T}\Pt{})^{\pm 1}$, where $a = (\Pt{}(0)^{T}
\Pt{}(0))^{-1}$. This follows from the fact that
\begin{equation}
  \E^{\int^t_0 G(t') \cos \theta(t') \; \D t'}
  = a' Q(t)^TQ(t),
\end{equation}
where $Q$ is any solution to Eq.~(\ref{eq:lin}), $\theta = \cov[Q]$,
and $a' = (Q(0)^T Q(0))^{-1}$. The above can be verified by a
straight-forward calculation of the logarithmic derivative of $Q^T Q$.
Specifically, for the unstable Floquet solution $\Fq{} = \exp(|\BZ|t)
\Pt{}$ we then have
\begin{equation}
  \E^{\int^t_0 G(t') \cos \theta(t') \; \D t'}
  = a \E^{2|\BZ|t} \Pt{}(t)^T\Pt{}(t),
\end{equation}
Using Eq.~(\ref{avarage_cos}) the claim follows.


\begin{thebibliography}{10}%
\makeatletter
\providecommand \@ifxundefined [1]{%
 \ifx #1\undefined \expandafter \@firstoftwo
 \else \expandafter \@secondoftwo
\fi
}%
\providecommand \@ifnum [1]{%
 \ifnum #1\expandafter \@firstoftwo
 \else \expandafter \@secondoftwo
\fi
}%
\providecommand \enquote [1]{``#1''}%
\providecommand \bibnamefont  [1]{#1}%
\providecommand \bibfnamefont [1]{#1}%
\providecommand \citenamefont [1]{#1}%
\providecommand\href[0]{\@sanitize\@href}%
\providecommand\@href[1]{\endgroup\@@startlink{#1}\endgroup\@@href}%
\providecommand\@@href[1]{#1\@@endlink}%
\providecommand \@sanitize [0]{\begingroup\catcode`\&12\catcode`\#12\relax}%
\@ifxundefined \pdfoutput {\@firstoftwo}{%
 \@ifnum{\z@=\pdfoutput}{\@firstoftwo}{\@secondoftwo}%
}{%
 \providecommand\@@startlink[1]{\leavevmode}%
 \providecommand\@@endlink[0]{}%
}{%
 \providecommand\@@startlink[1]{%
  \leavevmode
  \pdfstartlink
   attr{/Border[0 0 1 ]/H/I/C[0 1 1]}%
   user{/Subtype/Link/A<</Type/Action/S/URI/URI(#1)>>}%
  \relax
 }%
 \providecommand\@@endlink[0]{\pdfendlink}%
}%
\providecommand \url  [0]{\begingroup\@sanitize \@url }%
\providecommand \@url [1]{\endgroup\@href {#1}{\urlprefix}}%
\providecommand \urlprefix [0]{URL }%
\providecommand \Eprint[0]{\href }%
\@ifxundefined \urlstyle {%
  \providecommand \doi [1]{doi:\discretionary{}{}{}#1}%
}{%
  \providecommand \doi [0]{doi:\discretionary{}{}{}\begingroup
  \urlstyle{rm}\Url }%
}%
\providecommand \doibase [0]{http://dx.doi.org/}%
\providecommand \Doi[1]{\href{\doibase#1}}%
\providecommand \selectlanguage [0]{\@gobble}%
\providecommand \bibinfo [0]{\@secondoftwo}%
\providecommand \bibfield [0]{\@secondoftwo}%
\providecommand \translation [1]{[#1]}%
\providecommand \BibitemOpen[0]{}%
\providecommand \bibitemStop [0]{}%
\providecommand \bibitemNoStop [0]{.\EOS\space}%
\providecommand \EOS [0]{\spacefactor3000\relax}%
\providecommand \BibitemShut [1]{\csname bibitem#1\endcsname}%
\bibitem{sagdeev}%
  \BibitemOpen
  \bibfield{author}{%
  \bibinfo {author} {\bibfnamefont{R.~Z.}\ \bibnamefont{Sagdeev}}, \bibinfo
  {author} {\bibfnamefont{D.~A.}\ \bibnamefont{Usikov}},\ and\ \bibinfo
  {author} {\bibfnamefont{G.~M.}\ \bibnamefont{Zaslavsky}},\ }%
  \emph{\bibinfo {title} {Nonlinear Physics: From the Pendulum to Turbulence
  and Chaos}}\ (\bibinfo {publisher} {Harwood Academic Publishers},\ \bibinfo
  {address} {Chur, Switzerland},\ \bibinfo {year} {1992})\BibitemShut{NoStop}%
\bibitem{mccumber}%
  \BibitemOpen
  \bibfield{author}{%
  \bibinfo {author} {\bibfnamefont{D.~E.}\ \bibnamefont{McCumber}},\ }%
  \bibfield{title}{%
  \enquote{\bibinfo {title} {Effect of ac impedance on dc voltage-current
  characteristics of superconductor weak-link junctions},}\ }%
  \bibfield{journal}{%
  \bibinfo {journal} {J. Appl. Phys.}\ }%
  \textbf{\bibinfo {volume} {39}},\ \bibinfo {pages} {3113} (\bibinfo {year}
  {1968})\BibitemShut{NoStop}%
\bibitem{stewrt}%
  \BibitemOpen
  \bibfield{author}{%
  \bibinfo {author} {\bibfnamefont{W.~C.}\ \bibnamefont{Stewart}},\ }%
  \bibfield{title}{%
  \enquote{\bibinfo {title} {Current-voltage characteristics of {Josephson}
  junctions},}\ }%
  \bibfield{journal}{%
  \bibinfo {journal} {Appl. Phys. Lett.}\ }%
  \textbf{\bibinfo {volume} {12}},\ \bibinfo {pages} {277} (\bibinfo {year}
  {1968})\BibitemShut{NoStop}%
\bibitem{aslamazov69}%
  \BibitemOpen
  \bibfield{author}{%
  \bibinfo {author} {\bibfnamefont{L.~G.}\ \bibnamefont{Aslamazov}}\ and\
  \bibinfo {author} {\bibfnamefont{A.~I.}\ \bibnamefont{Larkin}},\ }%
  \bibfield{title}{%
  \enquote{\bibinfo {title} {Josephson effect in superconducting point
  contacts},}\ }%
  \bibfield{journal}{%
  \bibinfo {journal} {JETP Lett.}\ }%
  \textbf{\bibinfo {volume} {9}},\ \bibinfo {pages} {150} (\bibinfo {year}
  {1969})\BibitemShut{NoStop}%
\bibitem{langenberg}%
  \BibitemOpen
  \bibfield{author}{%
  \bibinfo {author} {\bibfnamefont{D.~N.}\ \bibnamefont{Langenberg}}, \bibinfo
  {author} {\bibfnamefont{D.~J.}\ \bibnamefont{Scalapino}}, \bibinfo {author}
  {\bibfnamefont{B.~N.}\ \bibnamefont{Taylor}},\ and\ \bibinfo {author}
  {\bibfnamefont{R.~E.}\ \bibnamefont{Eck}},\ }%
  \bibfield{title}{%
  \enquote{\bibinfo {title} {Microwave-induced dc voltages across {Josephson}
  junctions},}\ }%
  \bibfield{journal}{%
  \bibinfo {journal} {Phys. Lett.}\ }%
  \textbf{\bibinfo {volume} {20}},\ \bibinfo {pages} {563} (\bibinfo {year}
  {1966})\BibitemShut{NoStop}%
\bibitem{levinsen}%
  \BibitemOpen
  \bibfield{author}{%
  \bibinfo {author} {\bibfnamefont{M.~T.}\ \bibnamefont{Levinsen}}, \bibinfo
  {author} {\bibfnamefont{R.~Y.}\ \bibnamefont{Chiao}}, \bibinfo {author}
  {\bibfnamefont{M.~J.}\ \bibnamefont{Feldman}},\ and\ \bibinfo {author}
  {\bibfnamefont{B.~A.}\ \bibnamefont{Tucker}},\ }%
  \bibfield{title}{%
  \enquote{\bibinfo {title} {An inverse ac {Josephson} effect voltage
  standard},}\ }%
  \bibfield{journal}{%
  \bibinfo {journal} {Appl. Phys. Lett.}\ }%
  \textbf{\bibinfo {volume} {31}},\ \bibinfo {pages} {776} (\bibinfo {month}
  {12}\ \bibinfo {year} {1977})\BibitemShut{NoStop}%
\bibitem{hamilton}%
  \BibitemOpen
  \bibfield{author}{%
  \bibinfo {author} {\bibfnamefont{C.~A.}\ \bibnamefont{Hamilton}},\ }%
  \bibfield{title}{%
  \enquote{\bibinfo {title} {{Josephson} voltage standards},}\ }%
  \bibfield{journal}{%
  \bibinfo {journal} {Rev. Sci. Instr.}\ }%
  \textbf{\bibinfo {volume} {71}},\ \bibinfo {pages} {3611} (\bibinfo {year}
  {2000})\BibitemShut{NoStop}%
\bibitem{alekseev98:spont-dc}%
  \BibitemOpen
  \bibfield{author}{%
  \bibinfo {author} {\bibfnamefont{K.~N.}\ \bibnamefont{Alekseev}}, \bibinfo
  {author} {\bibfnamefont{E.~H.}\ \bibnamefont{Cannon}}, \bibinfo {author}
  {\bibfnamefont{J.~C.}\ \bibnamefont{McKinney}}, \bibinfo {author}
  {\bibfnamefont{F.~V.}\ \bibnamefont{Kusmartsev}},\ and\ \bibinfo {author}
  {\bibfnamefont{D.~K.}\ \bibnamefont{Campbell}},\ }%
  \bibfield{title}{%
  \enquote{\bibinfo {title} {Spontaneous dc current generation in a resistively
  shunted semiconductor superlattice driven by a terahertz field},}\ }%
  \bibfield{journal}{%
  \bibinfo {journal} {Phys. Rev. Lett.}\ }%
  \textbf{\bibinfo {volume} {80}},\ \bibinfo {pages} {2669} (\bibinfo {year}
  {1998})\BibitemShut{NoStop}%
\bibitem{cannon01}%
  \BibitemOpen
  \bibfield{author}{%
  \bibinfo {author} {\bibfnamefont{K.~N.}\ \bibnamefont{Alekseev}}, \bibinfo
  {author} {\bibfnamefont{E.~H.}\ \bibnamefont{Cannon}}, \bibinfo {author}
  {\bibfnamefont{F.~V.}\ \bibnamefont{Kusmartsev}},\ and\ \bibinfo {author}
  {\bibfnamefont{D.~K.}\ \bibnamefont{Campbell}},\ }%
  \bibfield{title}{%
  \enquote{\bibinfo {title} {Fractional and unquantized dc voltage generation
  in {THz}-driven semiconductor superlattices},}\ }%
  \bibfield{journal}{%
  \bibinfo {journal} {Europhys. Lett.}\ }%
  \textbf{\bibinfo {volume} {56}},\ \bibinfo {pages} {842} (\bibinfo {year}
  {2001}),\ \bibinfo {note} {[Erratum, K. N. Alekseev \textit{et al.} Europhys.
  Lett. \textbf{68}, 753 (2004)]}\BibitemShut{NoStop}%
\bibitem{alekseev02a}%
  \BibitemOpen
  \bibfield{author}{%
  \bibinfo {author} {\bibfnamefont{K.~N.}\ \bibnamefont{Alekseev}}\ and\
  \bibinfo {author} {\bibfnamefont{F.~V.}\ \bibnamefont{Kusmartsev}},\ }%
  \bibfield{title}{%
  \enquote{\bibinfo {title} {Pendulum limit, chaos and phase-locking in the
  dynamics of ac-driven semiconductor superlattices},}\ }%
  \bibfield{journal}{%
  \bibinfo {journal} {Phys. Lett. A}\ }%
  \textbf{\bibinfo {volume} {305}},\ \bibinfo {pages} {281} (\bibinfo {year}
  {2002})\BibitemShut{NoStop}%
\bibitem{dodin98b}%
  \BibitemOpen
  \bibfield{author}{%
  \bibinfo {author} {\bibfnamefont{E.~P.}\ \bibnamefont{Dodin}}, \bibinfo
  {author} {\bibfnamefont{A.~A.}\ \bibnamefont{Zharov}},\ and\ \bibinfo
  {author} {\bibfnamefont{A.~A.}\ \bibnamefont{Ignatov}},\ }%
  \bibfield{title}{%
  \enquote{\bibinfo {title} {Lateral superlattices in a strong electromagnetic
  field: self-induced transparency, multistability, and frequency
  multiplication},}\ }%
  \bibfield{journal}{%
  \bibinfo {journal} {J. Exp. Theor. Phys}\ }%
  \textbf{\bibinfo {volume} {87}},\ \bibinfo {pages} {1226} (\bibinfo {year}
  {1998})\BibitemShut{NoStop}%
\bibitem{alekseev05:lsslcr}%
  \BibitemOpen
  \bibfield{author}{%
  \bibinfo {author} {\bibfnamefont{K.~N.}\ \bibnamefont{Alekseev}}, \bibinfo
  {author} {\bibfnamefont{P.}~\bibnamefont{Pietiläinen}}, \bibinfo {author}
  {\bibfnamefont{J.}~\bibnamefont{Isoh\"at\"al\"a}}, \bibinfo {author}
  {\bibfnamefont{A.~A.}\ \bibnamefont{Zharov}},\ and\ \bibinfo {author}
  {\bibfnamefont{F.~V.}\ \bibnamefont{Kusmartsev}},\ }%
  \bibfield{title}{%
  \enquote{\bibinfo {title} {Chaos and rectification of electromagnetic wave in
  a lateral semiconductor superlattice},}\ }%
  \bibfield{journal}{%
  \bibinfo {journal} {Europhys. Lett.}\ }%
  \textbf{\bibinfo {volume} {70}},\ \bibinfo {pages} {292} (\bibinfo {year}
  {2005})\BibitemShut{NoStop}%
\bibitem{zharov06}%
  \BibitemOpen
  \bibfield{author}{%
  \bibinfo {author} {\bibfnamefont{A.~A.}\ \bibnamefont{Zharov}}\ and\ \bibinfo
  {author} {\bibfnamefont{A.~M.}\ \bibnamefont{Malkin}},\ }%
  \bibfield{title}{%
  \enquote{\bibinfo {title} {Generation of unipolar pulses during interaction
  between electromagnetic radiation and a lateral semiconductor
  superlattice},}\ }%
  \bibfield{journal}{%
  \bibinfo {journal} {Radiophys. Quantum Electron.}\ }%
  \textbf{\bibinfo {volume} {49}},\ \bibinfo {pages} {203} (\bibinfo {year}
  {2006})\BibitemShut{NoStop}%
\bibitem{dodin03}%
  \BibitemOpen
  \bibfield{author}{%
  \bibinfo {author} {\bibfnamefont{E.~P.}\ \bibnamefont{Dodin}}, \bibinfo
  {author} {\bibfnamefont{A.~A.}\ \bibnamefont{Zharov}},\ and\ \bibinfo
  {author} {\bibfnamefont{A.~M.}\ \bibnamefont{Malkin}},\ }%
  \bibfield{title}{%
  \enquote{\bibinfo {title} {Excitation of {Bloch} oscillations in a lateral
  semiconductor superlattice under the influence of electromagnetic pulses},}\
  }%
  \bibfield{journal}{%
  \bibinfo {journal} {J. Exp. Theor. Phys.}\ }%
  \textbf{\bibinfo {volume} {99}},\ \bibinfo {pages} {552} (\bibinfo {year}
  {2004})\BibitemShut{NoStop}%
\bibitem{pikovsky01}%
  \BibitemOpen
  \bibfield{author}{%
  \bibinfo {author} {\bibfnamefont{A.}~\bibnamefont{Pikovsky}}, \bibinfo
  {author} {\bibfnamefont{M.}~\bibnamefont{Rosenblum}},\ and\ \bibinfo {author}
  {\bibfnamefont{J.}~\bibnamefont{Kurths}},\ }%
  \emph{\bibinfo {title} {Synchronization}}\ (\bibinfo {publisher} {Cambridge
  University Press},\ \bibinfo {year} {2001})\BibitemShut{NoStop}%
\bibitem{gaifullin08}%
  \BibitemOpen
  \bibfield{author}{%
  \bibinfo {author} {\bibfnamefont{M.~B.}\ \bibnamefont{Gaifullin}}, \bibinfo
  {author} {\bibfnamefont{K.}~\bibnamefont{Hirata}}, \bibinfo {author}
  {\bibfnamefont{S.}~\bibnamefont{Ooi}}, \bibinfo {author}
  {\bibfnamefont{S.}~\bibnamefont{Savel'ev}}, \bibinfo {author}
  {\bibfnamefont{Yu.~I.}\ \bibnamefont{Latyshev}},\ and\ \bibinfo {author}
  {\bibfnamefont{T}~\bibnamefont{Mochiku}},\ }%
  \bibfield{title}{%
  \enquote{\bibinfo {title} {Synchronization in stacked array of the
  {Josephson} junctions in {Bi$_2$Sr$_2$CaCu$_2$O$_{8+\delta}$}},}\ }%
  \bibfield{journal}{%
  \bibinfo {journal} {Physica C}\ }%
  \textbf{\bibinfo {volume} {468}},\ \bibinfo {pages} {1896} (\bibinfo {year}
  {2008})\BibitemShut{NoStop}%
\bibitem{gaifullin09}%
  \BibitemOpen
  \bibfield{author}{%
  \bibinfo {author} {\bibfnamefont{V.~N.}\ \bibnamefont{Pavlenko}}, \bibinfo
  {author} {\bibfnamefont{Yu.~I.}\ \bibnamefont{Latyshev}}, \bibinfo {author}
  {\bibfnamefont{J.}~\bibnamefont{Chen}}, \bibinfo {author}
  {\bibfnamefont{M.~B.}\ \bibnamefont{Gaifullin}}, \bibinfo {author}
  {\bibfnamefont{A.}~\bibnamefont{Irzhak}}, \bibinfo {author}
  {\bibfnamefont{S.-J.}\ \bibnamefont{Kim}},\ and\ \bibinfo {author}
  {\bibfnamefont{P.~H.}\ \bibnamefont{Wu}},\ }%
  \bibfield{title}{%
  \enquote{\bibinfo {title} {Collective responses of {Bi-2212} stacked junction
  to 100 {GHz} microwave radiation under magnetic field oriented along the
  c-axis},}\ }%
  \bibfield{journal}{%
  \bibinfo {journal} {JETP Lett.}\ }%
  \textbf{\bibinfo {volume} {89}},\ \bibinfo {pages} {249} (\bibinfo {year}
  {2009})\BibitemShut{NoStop}%
\bibitem{kuzmin79}%
  \BibitemOpen
  \bibfield{author}{%
  \bibinfo {author} {\bibfnamefont{L.~S.}\ \bibnamefont{Kuzmin}}, \bibinfo
  {author} {\bibfnamefont{K.~K.}\ \bibnamefont{Likharev}},\ and\ \bibinfo
  {author} {\bibfnamefont{V.~V.}\ \bibnamefont{Migulin}},\ }%
  \bibfield{title}{%
  \enquote{\bibinfo {title} {Properties of parametric amplifiers using
  {Josephson} junctions with external pumping},}\ }%
  \bibfield{journal}{%
  \bibinfo {journal} {IEEE Trans. Magn.}\ }%
  \textbf{\bibinfo {volume} {15}},\ \bibinfo {pages} {454} (\bibinfo {year}
  {1979})\BibitemShut{NoStop}%
\bibitem{kuzmin80}%
  \BibitemOpen
  \bibfield{author}{%
  \bibinfo {author} {\bibfnamefont{L.~S.}\ \bibnamefont{Kuz'min}}, \bibinfo
  {author} {\bibfnamefont{K.~K.}\ \bibnamefont{Likharev}},\ and\ \bibinfo
  {author} {\bibfnamefont{V.~V.}\ \bibnamefont{Migulin}},\ }%
  \bibfield{title}{%
  \enquote{\bibinfo {title} {Properties of a one-frequency externally pumped
  nondegenerate {Josephson} contact parametric amplifier},}\ }%
  \bibfield{journal}{%
  \bibinfo {journal} {Radio Engineering and Electronic Physics}\ }%
  \textbf{\bibinfo {volume} {25}},\ \bibinfo {pages} {108} (\bibinfo {year}
  {1980})\BibitemShut{NoStop}%
\bibitem{kuzmin80-orig}%
  \BibitemOpen
  \bibfield{author}{%
  \bibinfo {author} {\bibfnamefont{L.~S.}\ \bibnamefont{Kuz'min}}, \bibinfo
  {author} {\bibfnamefont{K.~K.}\ \bibnamefont{Likharev}},\ and\ \bibinfo
  {author} {\bibfnamefont{V.~V.}\ \bibnamefont{Migulin}},\ }%
  \bibfield{journal}{%
  \bibinfo {journal} {Radiotekhnika-i-Elektronika}\ }%
  \textbf{\bibinfo {volume} {25}},\ \bibinfo {pages} {2195} (\bibinfo {year}
  {1980})\BibitemShut{NoStop}%
\bibitem{velichko99}%
  \BibitemOpen
  \bibfield{author}{%
  \bibinfo {author} {\bibfnamefont{A.~V.}\ \bibnamefont{Velichko}}\ and\
  \bibinfo {author} {\bibfnamefont{A.}~\bibnamefont{Porch}},\ }%
  \bibfield{title}{%
  \enquote{\bibinfo {title} {Modelling the nonlinear high-frequency response of
  a short {Josephson} junction under two-frequency irradiation},}\ }%
  \bibfield{journal}{%
  \bibinfo {journal} {IEEE Trans. Appl. Supercond.}\ }%
  \textbf{\bibinfo {volume} {9}},\ \bibinfo {pages} {2133} (\bibinfo {year}
  {1999})\BibitemShut{NoStop}%
\bibitem{likharev-c10-11}%
  \BibitemOpen
  \bibfield{author}{%
  \bibinfo {author} {\bibfnamefont{K.}~\bibnamefont{Likharev}},\ }%
  \enquote{\bibinfo {title} {Dynamics of {Josephson} junctions and circuits},}\
  \ (\bibinfo {publisher} {Gordon and Breach},\ \bibinfo {address} {New York},\
  \bibinfo {year} {1986})\ Chap.\ \bibinfo {chapter} {10 and
  11}\BibitemShut{NoStop}%
\bibitem{isohatala05:sbpend}%
  \BibitemOpen
  \bibfield{author}{%
  \bibinfo {author} {\bibfnamefont{J.}~\bibnamefont{Isoh\"at\"al\"a}}, \bibinfo
  {author} {\bibfnamefont{K.~N.}\ \bibnamefont{Alekseev}}, \bibinfo {author}
  {\bibfnamefont{L.~T.}\ \bibnamefont{Kurki}},\ and\ \bibinfo {author}
  {\bibfnamefont{P.}~\bibnamefont{Pietil\"ainen}},\ }%
  \bibfield{title}{%
  \enquote{\bibinfo {title} {Symmetry breaking in driven and strongly damped
  pendulum},}\ }%
  \bibfield{journal}{%
  \bibinfo {journal} {Phys. Rev. E}\ }%
  \textbf{\bibinfo {volume} {71}},\ \bibinfo {pages} {066206} (\bibinfo {year}
  {2005})\BibitemShut{NoStop}%
\bibitem{romeiras87}%
  \BibitemOpen
  \bibfield{author}{%
  \bibinfo {author} {\bibfnamefont{Felipe~J.}\ \bibnamefont{Romeiras}},
  \bibinfo {author} {\bibfnamefont{Anders}\ \bibnamefont{Bondeson}}, \bibinfo
  {author} {\bibfnamefont{Edward}\ \bibnamefont{Ott}}, \bibinfo {author}
  {\bibfnamefont{Thomas~M.}\ \bibnamefont{Antonsen}, \bibfnamefont{Jr}},\ and\
  \bibinfo {author} {\bibfnamefont{Celso}\ \bibnamefont{Grebogi}},\ }%
  \bibfield{title}{%
  \enquote{\bibinfo {title} {Quasiperiodically forced dynamical systems with
  strange nonchaotic attractors},}\ }%
  \bibfield{journal}{%
  \bibinfo {journal} {Physica D}\ }%
  \textbf{\bibinfo {volume} {26}},\ \bibinfo {pages} {277} (\bibinfo {year}
  {1987})\BibitemShut{NoStop}%
\bibitem{bumyalene89}%
  \BibitemOpen
  \bibfield{author}{%
  \bibinfo {author} {\bibfnamefont{S.}~\bibnamefont{Bumyalene}}, \bibinfo
  {author} {\bibfnamefont{G.}~\bibnamefont{Lasene}},\ and\ \bibinfo {author}
  {\bibfnamefont{K.}~\bibnamefont{Piragas}},\ }%
  \bibfield{title}{%
  \enquote{\bibinfo {title} {Rectification of an alternating current and
  generation of even harmonics in homogeneous semiconductors with an
  antisymmetric current-voltage characteristic},}\ }%
  \bibfield{journal}{%
  \bibinfo {journal} {Fiz. Tekh. Poluprovdn. (S.-Peterburg)}\ }%
  \textbf{\bibinfo {volume} {23}},\ \bibinfo {pages} {1479} (\bibinfo {year}
  {1989}),\ \bibinfo {note} {[Sov. Phys. Semicond. {\bf 23}, 918
  (1989)]}\BibitemShut{NoStop}%
\bibitem{dhumieres82}%
  \BibitemOpen
  \bibfield{author}{%
  \bibinfo {author} {\bibfnamefont{D.}~\bibnamefont{D'Humieres}}, \bibinfo
  {author} {\bibfnamefont{M.~R.}\ \bibnamefont{Beasley}}, \bibinfo {author}
  {\bibfnamefont{B.~A.}\ \bibnamefont{Huberman}},\ and\ \bibinfo {author}
  {\bibfnamefont{A.}~\bibnamefont{Libchaber}},\ }%
  \bibfield{title}{%
  \enquote{\bibinfo {title} {Chaotic states and routes to chaos in the forced
  pendulum},}\ }%
  \bibfield{journal}{%
  \bibinfo {journal} {Phys. Rev. A}\ }%
  \textbf{\bibinfo {volume} {26}},\ \bibinfo {pages} {3483} (\bibinfo {year}
  {1982})\BibitemShut{NoStop}%
\bibitem{swift84}%
  \BibitemOpen
  \bibfield{author}{%
  \bibinfo {author} {\bibfnamefont{J.~W.}\ \bibnamefont{Swift}}\ and\ \bibinfo
  {author} {\bibfnamefont{K.}~\bibnamefont{Wiesenfeld}},\ }%
  \bibfield{title}{%
  \enquote{\bibinfo {title} {Suppression of period doubling in symmetric
  systems},}\ }%
  \bibfield{journal}{%
  \bibinfo {journal} {Phys. Rev. Lett.}\ }%
  \textbf{\bibinfo {volume} {52}},\ \bibinfo {pages} {705} (\bibinfo {year}
  {1984})\BibitemShut{NoStop}%
\bibitem{prufer26}%
  \BibitemOpen
  \bibfield{author}{%
  \bibinfo {author} {\bibfnamefont{H.}~\bibnamefont{Pr\"{u}fer}},\ }%
  \bibfield{title}{%
  \enquote{\bibinfo {title} {Neue {Herleitung} der {Sturm-Liouvilleschen}
  {Reihenentwicklung} stetiger {Funktionen}},}\ }%
  \bibfield{journal}{%
  \bibinfo {journal} {Math. Ann.}\ }%
  \textbf{\bibinfo {volume} {95}},\ \bibinfo {pages} {499} (\bibinfo {year}
  {1926})\BibitemShut{NoStop}%
\bibitem{bondeson85:quasi-pend}%
  \BibitemOpen
  \bibfield{author}{%
  \bibinfo {author} {\bibfnamefont{A.}~\bibnamefont{Bondeson}}, \bibinfo
  {author} {\bibfnamefont{E.}~\bibnamefont{Ott}},\ and\ \bibinfo {author}
  {\bibfnamefont{T.~M.~Jr.}\ \bibnamefont{Antonsen}},\ }%
  \bibfield{title}{%
  \enquote{\bibinfo {title} {Quasiperiodically forced damped pendula and
  {Schrödinger} equations with quasiperiodic potentials: Implications of their
  equivalence},}\ }%
  \bibfield{journal}{%
  \bibinfo {journal} {Phys. Rev. Lett.}\ }%
  \textbf{\bibinfo {volume} {55}},\ \bibinfo {pages} {2103} (\bibinfo {year}
  {1985})\BibitemShut{NoStop}%
\bibitem{flach00}%
  \BibitemOpen
  \bibfield{author}{%
  \bibinfo {author} {\bibfnamefont{S.}~\bibnamefont{Flach}}, \bibinfo {author}
  {\bibfnamefont{O.}~\bibnamefont{Yevtushenko}},\ and\ \bibinfo {author}
  {\bibfnamefont{Y.}~\bibnamefont{Zolotaryuk}},\ }%
  \bibfield{title}{%
  \enquote{\bibinfo {title} {Directed current due to broken time-space
  symmetry},}\ }%
  \bibfield{journal}{%
  \bibinfo {journal} {Phys. Rev. Lett.}\ }%
  \textbf{\bibinfo {volume} {84}},\ \bibinfo {pages} {2358} (\bibinfo {year}
  {2000})\BibitemShut{NoStop}%
\bibitem{wiesenfeld85}%
  \BibitemOpen
  \bibfield{author}{%
  \bibinfo {author} {\bibfnamefont{K.}~\bibnamefont{Wiesenfeld}}\ and\ \bibinfo
  {author} {\bibfnamefont{B.}~\bibnamefont{McNamara}},\ }%
  \bibfield{title}{%
  \enquote{\bibinfo {title} {Period-doubling systems as small-signal
  amplifiers},}\ }%
  \bibfield{journal}{%
  \bibinfo {journal} {Phys. Rev. Lett.}\ }%
  \textbf{\bibinfo {volume} {55}},\ \bibinfo {pages} {13} (\bibinfo {year}
  {1985})\BibitemShut{NoStop}%
\bibitem{wiesenfeld86}%
  \BibitemOpen
  \bibfield{author}{%
  \bibinfo {author} {\bibfnamefont{K.}~\bibnamefont{Wiesenfeld}}\ and\ \bibinfo
  {author} {\bibfnamefont{B.}~\bibnamefont{McNamara}},\ }%
  \bibfield{title}{%
  \enquote{\bibinfo {title} {Small-signal amplification in bifurcating
  dynamical systems},}\ }%
  \bibfield{journal}{%
  \bibinfo {journal} {Phys. Rev. A}\ }%
  \textbf{\bibinfo {volume} {33}},\ \bibinfo {pages} {629} (\bibinfo {year}
  {1986})\BibitemShut{NoStop}%
\bibitem{johnson82}%
  \BibitemOpen
  \bibfield{author}{%
  \bibinfo {author} {\bibfnamefont{R.}~\bibnamefont{Johnson}}\ and\ \bibinfo
  {author} {\bibfnamefont{J.}~\bibnamefont{Moser}},\ }%
  \bibfield{title}{%
  \enquote{\bibinfo {title} {The rotation number for almost periodic
  potentials},}\ }%
  \bibfield{journal}{%
  \bibinfo {journal} {Commun. Math. Phys.}\ }%
  \textbf{\bibinfo {volume} {84}},\ \bibinfo {pages} {403} (\bibinfo {year}
  {1982})\BibitemShut{NoStop}%
\bibitem{kuznetsov98}%
  \BibitemOpen
  \bibfield{author}{%
  \bibinfo {author} {\bibfnamefont{Y.}~\bibnamefont{Kuznetsov}},\ }%
  \emph{\bibinfo {title} {Elements of Applied Bifurcation theory}}\ (\bibinfo
  {publisher} {Springer-Verlag},\ \bibinfo {address} {New {York}},\ \bibinfo
  {year} {1998})\BibitemShut{NoStop}%
\bibitem{isohatala09_footnote2}%
  \BibitemOpen
  \bibinfo {note} {This is the determinant of a matrix whose columns are the
  Floquet solutions. Fixing the value of this determinant at $t = 0$ fixes it
  for all $t$, because of the vanishing trace of $A$}\BibitemShut{NoStop}%
\bibitem{verhulst05}%
  \BibitemOpen
  \bibfield{author}{%
  \bibinfo {author} {\bibfnamefont{F.}~\bibnamefont{Verhulst}},\ }%
  \emph{\bibinfo {title} {Methods and Applications of Singular Perturbations}}\
  (\bibinfo {publisher} {Springer},\ \bibinfo {year}
  {2005})\BibitemShut{NoStop}%
\bibitem{murray84:asympanal}%
  \BibitemOpen
  \bibfield{author}{%
  \bibinfo {author} {\bibfnamefont{J.~D.}\ \bibnamefont{Murray}},\ }%
  \emph{\bibinfo {title} {Asymptotic Analysis}}\ (\bibinfo {publisher}
  {Springer-Verlag},\ \bibinfo {address} {New York},\ \bibinfo {year}
  {1984})\BibitemShut{NoStop}%
\bibitem{abramowitz}%
  \BibitemOpen
  \emph{\bibinfo {title} {Handbook of Mathematical Functions}},\ edited by\
  \bibinfo {editor} {\bibfnamefont{M.}~\bibnamefont{Abramowitz}}\ and\ \bibinfo
  {editor} {\bibfnamefont{I.}~\bibnamefont{Stegun}}\ (\bibinfo {publisher}
  {Dover},\ \bibinfo {address} {New York},\ \bibinfo {year}
  {1965})\BibitemShut{NoStop}%
\bibitem{chesca08}%
  \BibitemOpen
  \bibfield{author}{%
  \bibinfo {author} {\bibfnamefont{B.}~\bibnamefont{Chesca}}, \bibinfo {author}
  {\bibfnamefont{S.~E.}\ \bibnamefont{Savel'ev}}, \bibinfo {author}
  {\bibfnamefont{A.~L.}\ \bibnamefont{Rakhmanov}}, \bibinfo {author}
  {\bibfnamefont{H.~J.~H.}\ \bibnamefont{Smilde}},\ and\ \bibinfo {author}
  {\bibfnamefont{H.}~\bibnamefont{Hilgenkamp}},\ }%
  \bibfield{title}{%
  \enquote{\bibinfo {title} {Controlling {Josephson} dynamics by strong
  microwave fields},}\ }%
  \bibfield{journal}{%
  \bibinfo {journal} {Phys. Rev. B}\ }%
  \textbf{\bibinfo {volume} {78}},\ \bibinfo {pages} {094505} (\bibinfo {year}
  {2008})\BibitemShut{NoStop}%
\bibitem{ignatov76}%
  \BibitemOpen
  \bibfield{author}{%
  \bibinfo {author} {\bibfnamefont{A.~A.}\ \bibnamefont{Ignatov}}\ and\
  \bibinfo {author} {\bibfnamefont{Yu.~A.}\ \bibnamefont{Romanov}},\ }%
  \bibfield{title}{%
  \enquote{\bibinfo {title} {Nonlinear electromagnetic properties of
  semiconductors with superlattice},}\ }%
  \bibfield{journal}{%
  \bibinfo {journal} {Phys. Stat. Sol. B}\ }%
  \textbf{\bibinfo {volume} {73}},\ \bibinfo {pages} {327} (\bibinfo {year}
  {1976})\BibitemShut{NoStop}%
\bibitem{wacker02:ssreview}%
  \BibitemOpen
  \bibfield{author}{%
  \bibinfo {author} {\bibfnamefont{A.}~\bibnamefont{Wacker}},\ }%
  \bibfield{title}{%
  \enquote{\bibinfo {title} {Semiconductor superlattices: a model system for
  nonlinear transport},}\ }%
  \bibfield{journal}{%
  \bibinfo {journal} {Phys. Rep.}\ }%
  \textbf{\bibinfo {volume} {357}},\ \bibinfo {pages} {1} (\bibinfo {year}
  {2002})\BibitemShut{NoStop}%
\bibitem{alekseev96:dissp-chaos-ssl}%
  \BibitemOpen
  \bibfield{author}{%
  \bibinfo {author} {\bibfnamefont{K.~N.}\ \bibnamefont{Alekseev}}, \bibinfo
  {author} {\bibfnamefont{G.~P.}\ \bibnamefont{Berman}}, \bibinfo {author}
  {\bibfnamefont{D.~K.}\ \bibnamefont{Campbell}}, \bibinfo {author}
  {\bibfnamefont{E.~H.}\ \bibnamefont{Cannon}},\ and\ \bibinfo {author}
  {\bibfnamefont{M.~C.}\ \bibnamefont{Cargo}},\ }%
  \bibfield{title}{%
  \enquote{\bibinfo {title} {Dissipative chaos in semiconductor
  superlattices},}\ }%
  \bibfield{journal}{%
  \bibinfo {journal} {Phys. Rev. B}\ }%
  \textbf{\bibinfo {volume} {54}},\ \bibinfo {pages} {10625} (\bibinfo {year}
  {1996})\BibitemShut{NoStop}%
\bibitem{isohatala09_footnote1}%
  \BibitemOpen
  \bibinfo {note} {We cannot say anything about the stability of the solutions
  in our approximation, since the full second order equation should be
  considered. We can infer that the symmetry-broken regions are associated with
  a certain branch.}\BibitemShut{Stop}%
\bibitem{sekine05}%
  \BibitemOpen
  \bibfield{author}{%
  \bibinfo {author} {\bibfnamefont{N.}~\bibnamefont{Sekine}}\ and\ \bibinfo
  {author} {\bibfnamefont{K.}~\bibnamefont{Hirakawa}},\ }%
  \bibfield{title}{%
  \enquote{\bibinfo {title} {Dispersive terahertz gain of a nonclassical
  oscillator: {Bloch} oscillation in semiconductor superlattices},}\ }%
  \bibfield{journal}{%
  \bibinfo {journal} {Phys. Rev. Lett.}\ }%
  \textbf{\bibinfo {volume} {94}},\ \bibinfo {pages} {057408} (\bibinfo {year}
  {2005})\BibitemShut{NoStop}%
\bibitem{hyart09_magnetic}%
  \BibitemOpen
  \bibfield{author}{%
  \bibinfo {author} {\bibfnamefont{T.}~\bibnamefont{Hyart}}, \bibinfo {author}
  {\bibfnamefont{J.}~\bibnamefont{Mattas}},\ and\ \bibinfo {author}
  {\bibfnamefont{K.~N.}\ \bibnamefont{Alekseev}},\ }%
  \bibfield{title}{%
  \enquote{\bibinfo {title} {Model of the influence of an external magnetic
  field on the gain of terahertz radiation from semiconductor superlattices},}\
  }%
  \bibfield{journal}{%
  \bibinfo {journal} {Phys. Rev. Lett.}\ }%
  \textbf{\bibinfo {volume} {103}},\ \bibinfo {pages} {117401} (\bibinfo {year}
  {2009})\BibitemShut{NoStop}%
\bibitem{hyart08}%
  \BibitemOpen
  \bibfield{author}{%
  \bibinfo {author} {\bibfnamefont{T.}~\bibnamefont{Hyart}}, \bibinfo {author}
  {\bibfnamefont{K.~N.}\ \bibnamefont{Alekseev}},\ and\ \bibinfo {author}
  {\bibfnamefont{E.~V.}\ \bibnamefont{Thuneberg}},\ }%
  \bibfield{title}{%
  \enquote{\bibinfo {title} {Bloch gain in dc-ac-driven semiconductor
  superlattices in the absence of electric domains},}\ }%
  \bibfield{journal}{%
  \bibinfo {journal} {Phys. Rev. B}\ }%
  \textbf{\bibinfo {volume} {77}},\ \bibinfo {pages} {165330} (\bibinfo {year}
  {2008})\BibitemShut{NoStop}%
\bibitem{hyart09_quasistatic}%
  \BibitemOpen
  \bibfield{author}{%
  \bibinfo {author} {\bibfnamefont{T.}~\bibnamefont{Hyart}}, \bibinfo {author}
  {\bibfnamefont{N.~V.}\ \bibnamefont{Alexeeva}}, \bibinfo {author}
  {\bibfnamefont{J.}~\bibnamefont{Mattas}},\ and\ \bibinfo {author}
  {\bibfnamefont{K.~N.}\ \bibnamefont{Alekseev}},\ }%
  \bibfield{title}{%
  \enquote{\bibinfo {title} {Terahertz bloch oscillator with a modulated
  bias},}\ }%
  \bibfield{journal}{%
  \bibinfo {journal} {Phys. Rev. Lett.}\ }%
  \textbf{\bibinfo {volume} {102}},\ \bibinfo {pages} {140405} (\bibinfo {year}
  {2009})\BibitemShut{NoStop}%
\bibitem{minami09}%
  \BibitemOpen
  \bibfield{author}{%
  \bibinfo {author} {\bibfnamefont{H.}~\bibnamefont{Minami}}, \bibinfo {author}
  {\bibfnamefont{I.}~\bibnamefont{Kakeya}}, \bibinfo {author}
  {\bibfnamefont{H.}~\bibnamefont{Yamaguchi}}, \bibinfo {author}
  {\bibfnamefont{T.}~\bibnamefont{Yamamoto}},\ and\ \bibinfo {author}
  {\bibfnamefont{K.}~\bibnamefont{Kadowaki}},\ }%
  \bibfield{title}{%
  \enquote{\bibinfo {title} {Characteristics of terahertz radiation emitted
  from the intrinsic {Josephson} junctions in high-{$T_\text{c}$}
  superconductor {Bi$_2$Sr$_2$CaCu$_2$O$_{8+\delta}$}},}\ }%
  \bibfield{journal}{%
  \bibinfo {journal} {Appl. Phys. Lett.}\ }%
  \textbf{\bibinfo {volume} {95}},\ \bibinfo {pages} {232511} (\bibinfo {year}
  {2009})\BibitemShut{NoStop}%
\bibitem{song09}%
  \BibitemOpen
  \bibfield{author}{%
  \bibinfo {author} {\bibfnamefont{F.}~\bibnamefont{Song}}, \bibinfo {author}
  {\bibfnamefont{F.}~\bibnamefont{M\"uller}}, \bibinfo {author}
  {\bibfnamefont{R.}~\bibnamefont{Behr}},\ and\ \bibinfo {author}
  {\bibfnamefont{A.~M.}\ \bibnamefont{Klushin}},\ }%
  \bibfield{title}{%
  \enquote{\bibinfo {title} {Coherent emission from large arrays of discrete
  {Josephson} junctions},}\ }%
  \bibfield{journal}{%
  \bibinfo {journal} {Appl. Phys. Lett.}\ }%
  \textbf{\bibinfo {volume} {95}},\ \bibinfo {pages} {172501} (\bibinfo {year}
  {2009})\BibitemShut{NoStop}%
\end{thebibliography}
\end{document}